\documentclass[letterpaper]{article} 
\usepackage[preprint]{aaai2027}
\usepackage[hyphens]{url}  
\usepackage{graphicx} 
\usepackage{natbib}  
\usepackage{caption} 
\graphicspath{{.}}

\usepackage{booktabs}
\usepackage{amsfonts}
\usepackage{amsmath}
\usepackage{amssymb}
\usepackage{nicefrac}
\usepackage{microtype}
\usepackage{multirow}
\usepackage{xcolor}
\usepackage{subcaption}
\usepackage{algorithm}
\usepackage{algpseudocode}
\usepackage{tikz}
\usetikzlibrary{arrows.meta}
\usepackage{framed}  

\title{LASAR: Latent Adaptive Semantic Aligned Reasoning for Generative Recommendation}

\author{
    Yiwen Chen\textsuperscript{\rm 1,\rm 2},
    Fuwei Zhang\textsuperscript{\rm 1},
    Zehao Chen\textsuperscript{\rm 1},
    Hehan Li\textsuperscript{\rm 2},
    Peizhi Xu\textsuperscript{\rm 2},
    Hanmeng Liu\textsuperscript{\rm 2},
    Shuanglong Li\textsuperscript{\rm 2},
    Xin Pei\textsuperscript{\rm 2},
    Fuzhen Zhuang\textsuperscript{\rm 1}\thanks{Corresponding author: zhuangfuzhen@buaa.edu.cn},
    Zhao Zhang\textsuperscript{\rm 1}
}
\affiliations{
    \textsuperscript{\rm 1}School of Artificial Intelligence, Beihang University, Beijing, China\\
    \textsuperscript{\rm 2}Baidu, Beijing, China
}

\begin{document}

\maketitle

\begin{abstract}
Large Language Models (LLMs) have demonstrated powerful reasoning capabilities through Chain-of-Thought (CoT) in various tasks, yet the inefficiency of token-by-token generation hinders real-world deployment in latency-sensitive recommender systems. Latent reasoning has emerged as an effective paradigm in LLMs, performing multi-step inference in a continuous hidden-state space to achieve stronger reasoning at lower cost. However, this paradigm remains underexplored in mainstream generative recommendation. Achieving this reveals three key challenges: (1) the gap between prior-less Semantic ID (SID) symbols and continuous latent reasoning, as SIDs lack pre-trained semantics, hindering joint optimization; (2) representation drift due to a lack of reasoning chain supervision; and (3) the suboptimality of applying a globally fixed reasoning depth. To address these, we propose \textbf{LASAR} (Latent Adaptive Semantic Aligned Reasoning), an SFT-then-RL framework. \textbf{First}, we bridge this gap via two-stage training: Stage~1 grounds SID semantics before Stage~2 introduces latent reasoning, ensuring efficient convergence. \textbf{Second}, we mitigate representation drift through explicit CoT semantic alignment. Step-wise bidirectional KL divergence constrains the latent reasoning trajectory using hidden-state anchors extracted from CoT text, while a Policy Head predicts per-sample reasoning depth. \textbf{Third}, during the GRPO-based RL phase, terminal-only KL alignment accommodates variable-length reasoning, and REINFORCE optimizes the Policy Head to dynamically allocate steps. This nearly halves the average latent step count while simultaneously improving recommendation quality. Experiments on three real-world datasets show that LASAR achieves the best overall performance across the evaluated settings. It adds limited inference latency and is roughly 20$\times$ faster than generating explicit CoT text.
\end{abstract}

\section{Introduction}
\label{sec:intro}

\begin{figure}[t]
  \centering
  \includegraphics[width=0.95\linewidth]{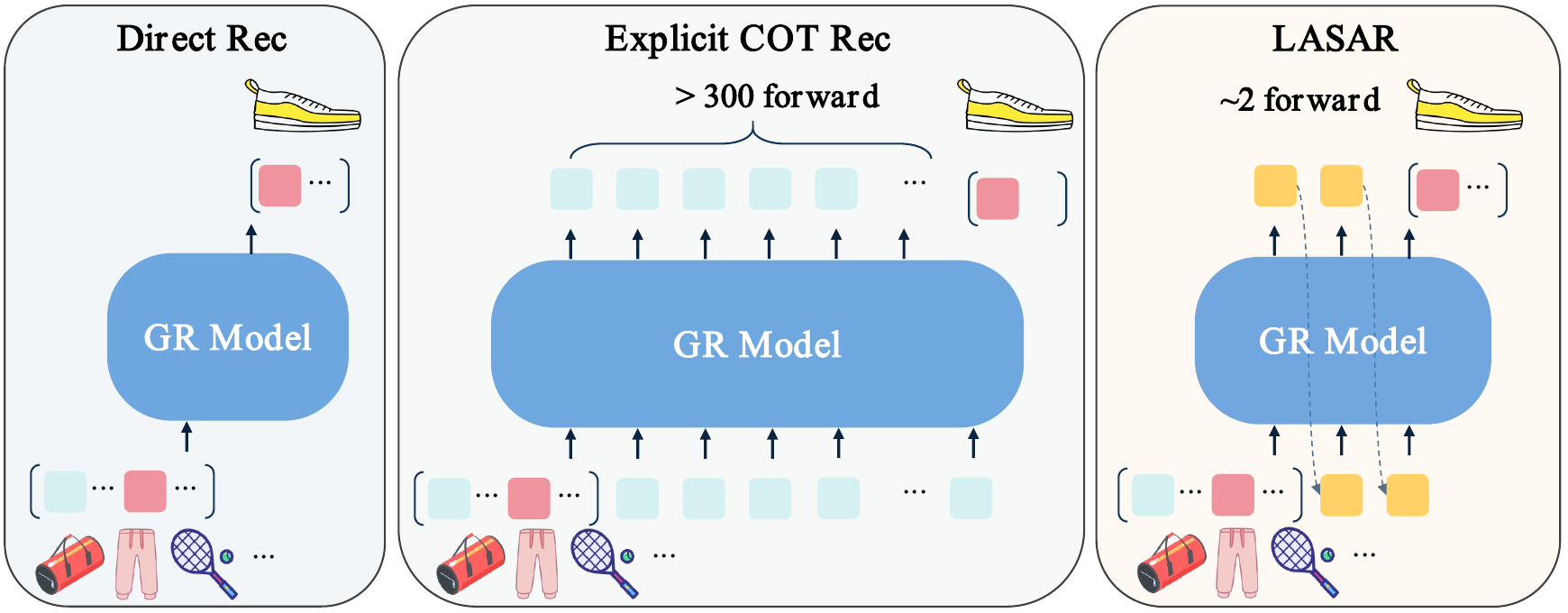}
  \vspace{-10pt}
  \caption{Motivation: long CoT chain to short latent loops.}
  \label{fig:motivation}
  \vspace{-24pt}
\end{figure}

Large Language Models (LLMs) in recommender systems are advancing rapidly along two paths. One is \textbf{generative recommendation}: P5~\cite{31_P5} and M6-Rec~\cite{32_M6Rec} pioneered the unified recommendation pretraining paradigm, TIGER~\cite{34_TIGER} introduced Semantic ID-based generative retrieval, LC-Rec~\cite{36_LCRec} integrated collaborative semantics into LLMs for direct item ID generation, while other studies~\cite{33_TALLRec,35_IndexItemIDs,38_E4SRec,40_EAGER} further advanced generative recommendation from alignment, indexing, and collaboration perspectives. MiniOneRec~\cite{37_MiniOneRec} built the first fully open-source generative recommendation framework. The other path is \textbf{LLM reasoning}: CoT~\cite{01_CoT,02_ZeroShot} improved task performance by explicitly generating intermediate reasoning steps, while subsequent studies~\cite{03_SelfConsistency,04_ToT,05_GoT,06_CoTTheory,07_ExpressivePower} extended the boundaries of explicit reasoning. However, even as DeepSeek-R1~\cite{22_DeepSeekR1} and O1~\cite{23_O1} have pushed explicit reasoning to its limits, the fundamental bottleneck of reasoning latency remains unresolved. To address this, Coconut~\cite{08_Coconut} proposed moving reasoning from the token space to the continuous latent space, realizing multi-step latent reasoning through a hidden state feedback loop, achieving stronger reasoning at lower cost. Subsequent studies~\cite{09_ImplicitCoT,10_FromExplicitCoTtoImplicitCoT,15_LoopedTransformers,16_Huginn,17_Heima,18_SoftCoT,19_LaTRO,20_ThinkAtHard,21_TokenAssorted} advanced this paradigm from various angles. Figure~\ref{fig:motivation} contrasts the resulting recommendation paradigms by their reasoning cost.

A natural question arises: \textbf{can such latent reasoning bring similar benefits to generative recommendation?} We survey existing work and find that the intersection remains surprisingly limited. In traditional discriminative recommendation, ReaRec~\cite{46_ReaRec}, LARES~\cite{47_LARES}, and other studies~\cite{48_STREAMRec,49_ManCAR,50_LCRSER} introduced latent reasoning, but they target the ID-embedding discriminative ranking paradigm, not the mainstream decoder-only generative paradigm. In generative recommendation, GREAM~\cite{41_GREAM} and other studies~\cite{42_RecSAVER,43_Reason4Rec,44_RecR1,45_R2EC} leverage explicit CoT reasoning, but text generation incurs high latency and introduces a fundamental tradeoff between reasoning-mode and direct-mode performance, and GREAM's own ablation reveals that its RL post-training (SRPO) degrades Direct recommendation metrics ($-5.3\%$ on Instruments), suggesting mode competition. Few preliminary studies~\cite{51_LatentR3,52_S2GR} attempt to introduce latent reasoning into LLM-based recommendation, but either perform only shallow single-step attention over history sequences or essentially insert tokens rather than performing recurrent iteration, so neither realizes the full Coconut-style hidden state feedback loop. To the best of our knowledge, \textbf{we are the first to perform complete latent reasoning through recurrent hidden-state feedback and adapt reasoning depth to each sample in mainstream generative recommendation.}

When introducing recurrent latent reasoning into a generative recommendation framework, we find that it is \textbf{not a ``free lunch''}: directly coupling SID grounding with latent reasoning degrades recommendation performance. We identify three key challenges in the generative recommendation setting: \textbf{(1) Semantic grounding gap between prior-less SIDs and latent reasoning.} Coconut works well in NLP because its tokens already carry rich pre-trained semantics. In generative recommendation, however, Semantic IDs (SIDs) are an entirely new symbolic system constructed from scratch with zero priors. Jointly training SID learning and latent reasoning forces the model to simultaneously ground a new symbol system \emph{and} perform continuous-space reasoning, causing optimization collapse: without semantic anchors, latent reasoning has no stable foundation to evolve on (Figure~\ref{fig:convergence}). \textbf{(2) Representation drift.} Recommendation lacks ground-truth reasoning chain supervision: there is no ``standard reasoning process'' to reference. Directly introducing latent reasoning without semantic constraints causes hidden states to drift in continuous space toward meaningless representations, and naively adding latent reasoning without alignment brings negligible gains (Table~\ref{tab:sft_ablation}). \textbf{(3) Inflexible and inefficient fixed-step reasoning.} Coconut and ReaRec both adopt a globally fixed number of reasoning steps $K$, applying the same reasoning depth to all samples. A fixed budget is fundamentally suboptimal: many samples can be correctly answered with minimal reasoning, while others benefit from deeper inference.

To address these challenges, we propose \textbf{LASAR} (Latent Adaptive
Semantic Aligned Reasoning), the first framework for adaptive
recurrent latent reasoning in mainstream generative recommendation.
We further engineer the complete training-and-inference pipeline, with
batched variable-depth computation, latent KV-cache reuse,
and trie-constrained beam decoding. LASAR adopts an SFT-then-RL paradigm
with three corresponding designs:

\textbf{(1) Two-stage decoupling bridges the semantic grounding gap.} Within the SFT phase, latent reasoning is deferred to Stage~2, after the model has grounded SID semantics and established basic recommendation capability, resolving the convergence bottleneck and achieving $\sim$3$\times$ faster convergence (Figure~\ref{fig:convergence}). \textbf{(2) Explicit CoT semantic alignment addresses representation drift.} We semantically segment explicit CoT reasoning text, extract hidden states as semantic anchors, and align each latent step to the corresponding CoT segment anchor via bidirectional KL divergence (Section~\ref{sec:sft}). The goal is not to replicate explicit reasoning, but to provide semantic guidance for the latent reasoning trajectory. \textbf{(3) Policy Head + REINFORCE enables flexible step optimization.} A Policy Head, warm-started during SFT, predicts reasoning depth per sample. Then during the GRPO-based RL phase, REINFORCE optimizes the Policy Head's step allocation, nearly halving the average step count while improving recommendation quality (Section~\ref{sec:step_opt}). At inference, LASAR incurs limited additional latency over MiniOneRec while remaining over 20$\times$ faster than explicit CoT (Table~\ref{tab:efficiency}).

Experiments on three real-world datasets show LASAR achieves best performance on nearly all metric-dataset combinations. Ablation reveals the failure modes of latent reasoning and confirms that all components make distinct and complementary contributions.

\vspace{-4pt}
\section{Methodology}
\label{sec:method}

\begin{figure*}[t]
  \vspace{-48pt}
  \centering
  \includegraphics[width=0.85\textwidth]{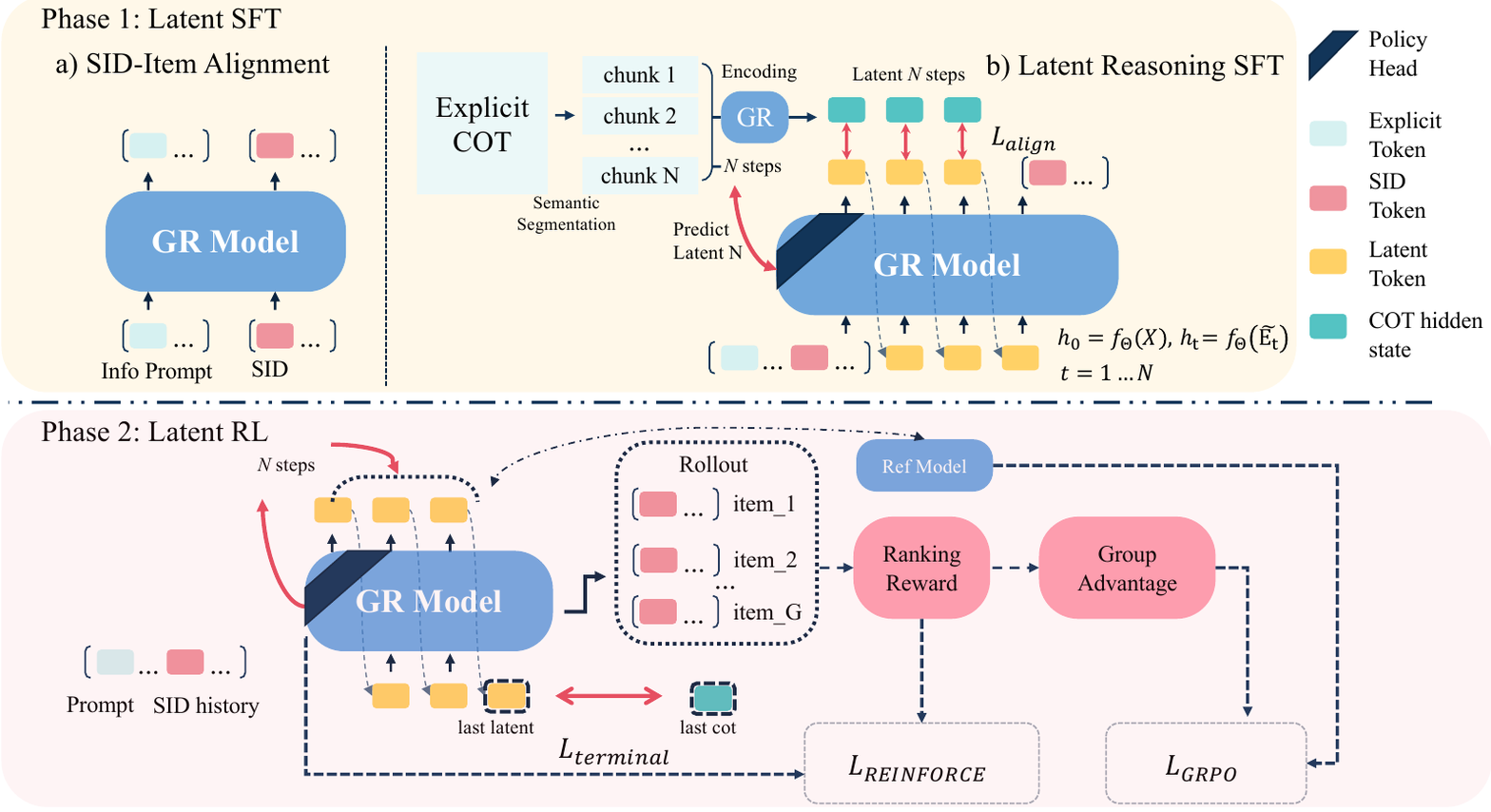}
  \vspace{-4em}
  \caption{LASAR framework overview. A hidden-state feedback loop iteratively refines latent tokens in continuous space, while a Policy Head predicts per-sample reasoning depth $N$ for adaptive reasoning. SFT: two-stage decoupling + step-wise bidirectional KL alignment with CoT anchors. RL: GRPO (generation quality) + REINFORCE (adaptive reasoning efficiency) + Terminal KL (semantic consistency).}
  \label{fig:architecture}
  \vspace{-15pt}
  \end{figure*}

The Introduction identified three challenges in adapting latent reasoning to generative recommendation. LASAR addresses them through three corresponding designs: a latent reasoning mechanism with sample-level adaptive step prediction (Section~\ref{sec:latent_mechanism}), an SFT phase that bridges the semantic grounding gap and representation drift via two-stage decoupling and CoT semantic alignment (Section~\ref{sec:sft}), and an RL phase that jointly optimizes generation quality and reasoning efficiency (Section~\ref{sec:rl}). We first define the generative recommendation task, then describe each component in turn.

\subsection{Problem Definition}

Let $\mathcal{I}$ be the set of items, where each item $i \in \mathcal{I}$ is associated with text features (e.g., title, description). Given a user's chronological interaction history $\mathcal{S} = \{i_1, i_2, \dots, i_t\}$, the objective of sequential recommendation is to predict the next item $i_{t+1}$ the user will interact with.

\textbf{Item Tokenization.}
To leverage the generative power of LLMs, each item $i$ is represented as a unique sequence of $M$ hierarchical discrete tokens, termed Semantic ID (SID). An item $i$ is mapped to its SID via a quantization function $Q(\cdot)$ applied to its text embedding $\mathbf{e}_i$:
{\small
\begin{equation}
\mathrm{SID}(i) = Q(\mathbf{e}_i) = (s_1, s_2, \dots, s_M), \quad s_j \in \mathcal{C}^{(j)},
\end{equation}

}
where $\mathcal{C}^{(j)}$ denotes the $j$-th codebook. We follow the Residual Quantization K-Means pipeline, which has been shown to be a strong choice in generative recommendation~\cite{41_GREAM,61_HandbookSID}. The resulting tokens are integrated into the LLM's vocabulary as special identifiers to capture collaborative signals.

\textbf{Generative Recommendation.}
In this framework, the recommendation task is reformulated as conditional sequence generation. We construct the input token sequence $X$ by combining a natural-language prompt ($\text{text}_{\text{nl}}$, e.g., task instructions, user or item textual descriptions) with the SID token sequences of the user's history $\mathcal{S}$:
$X = \big[\text{text}_{\text{nl}},\; \mathrm{SID}(i_1),\; \mathrm{SID}(i_2),\; \dots,\; \mathrm{SID}(i_t)\big]$,
where each $\mathrm{SID}(i_j) = (s_1, \dots, s_M)$ contributes $M$ special tokens. The model generates the target $\mathbf{Y} = \mathrm{SID}(i_{t+1})$ autoregressively:
{\small
\begin{equation}
p(\mathbf{Y} \mid X; \Theta) = \prod_{k=1}^{M} p(y_k \mid X, y_1, \dots, y_{k-1}; \Theta),
\end{equation}
}

where $y_k$ is the $k$-th token of the target SID and $\Theta$ denotes the backbone parameters. Our work primarily focuses on the architectural design of the backbone $\Theta$ to better capture the complex dependencies within $p(\mathbf{Y} \mid X)$.

\subsection{Latent Reasoning Mechanism}
\label{sec:latent_mechanism}

LASAR is built on a backbone LLM and performs multi-step reasoning in continuous latent space through a hidden-state feedback loop (Figure~\ref{fig:architecture}). The training follows an SFT-then-RL paradigm: the SFT phase (Section~\ref{sec:sft}) establishes latent reasoning capability with CoT semantic alignment, and the RL phase (Section~\ref{sec:rl}) jointly optimizes generation quality and reasoning efficiency.

\textbf{Latent Token Design.} Three special tokens, \texttt{<s>} (start), \texttt{<t>} (thought, repeated $N$ times), \texttt{<e>} (end), are inserted between the prompt and answer, forming \texttt{[Prompt] <s> <t>$\times$N <e> [Answer]}. Unlike Coconut and ReaRec's globally fixed step count $K$, LASAR predicts a sample-specific $N$ via a Policy Head (Section~\ref{sec:latent_policy}).

\textbf{Recurrent Latent Loop.}
The core mechanism is a hidden-state feedback loop. Let $h_0 \in \mathbb{R}^{D}$ denote the last-layer hidden state at the final prompt token. The latent reasoning process,
{\small
\begin{equation}
h_0 = f_\Theta(X), \qquad
h_t = f_\Theta\big(\tilde{E}_t\big), \quad t = 1, \dots, N,
\label{eq:latent_loop}
\end{equation}

}
where $\tilde{E}_t = [E_X,\; h_0,\; h_1,\; \dots,\; h_{t-1}]$ is the augmented input embedding sequence, where $E_X$ denotes the token embeddings of $X$. Each subsequent position replaces the standard token embedding with the previous step's hidden state $h_{t-1}$. After $N$ iterations, the answer segment is generated autoregressively starting from $h_N$, reusing the accumulated KV cache to avoid recomputing the prompt and latent steps.
This design realizes the Coconut-style continuous-space reasoning loop~\cite{08_Coconut}: intermediate states are unobservable dense vectors, and the model iteratively refines its reasoning without generating any discrete tokens.

\textbf{Adaptive Step Allocation via Policy Head.}\label{sec:policy_head}
\label{sec:latent_policy}
The Policy Head is a two-layer MLP that predicts the step count $N$ from the prompt-final hidden state $h_0$:
$\pi_\phi(\cdot \mid h_0) = \text{Softmax}\big(W_2 \cdot \tanh(W_1 \cdot h_0 + b_1) + b_2\big)$
with output dimension $N_{\text{max}}$ (maximum reasoning steps, default 8). During SFT, each sample's latent depth $N_i$ equals its number of CoT semantic segments (Section~\ref{sec:sft_align}), and the batch loop runs to $\max_i N_i$; the Policy Head is concurrently CE-trained to predict $N_i$ as a warm start. During RL (Section~\ref{sec:rl}), the Policy Head switches to sampling $N \sim \pi_\phi$ and is optimized via REINFORCE (Section~\ref{sec:reinforce}); at inference, $N = \text{argmax}(\pi_\phi)$. A key advantage of pre-predicting $N$ before the latent loop is that all beams of the same prompt share the same $N$, making the computation graph fully determined during rollout and simplifying batch beam search.

\begin{figure}[h]
  \vspace{-0.5em}

  \centering
  \resizebox{0.80\linewidth}{!}{%
  \begin{tabular}{c | c c c c c c c | c}
  \toprule
  & \multicolumn{7}{c|}{Latent region} & \\
  Sample & \texttt{<s>} & \texttt{<t>} & \texttt{<t>} & \texttt{<t>} & \texttt{<t>} & \texttt{<t>} & \texttt{<e>} & Answer \\
  \midrule
  A ($N{=}2$) & \texttt{<s>} & \texttt{<t>} & \texttt{<t>} & \texttt{PAD} & \texttt{PAD} & \texttt{PAD} & \texttt{<e>} & \texttt{[ans]} \\
  B ($N{=}3$) & \texttt{<s>} & \texttt{<t>} & \texttt{<t>} & \texttt{<t>} & \texttt{PAD} & \texttt{PAD} & \texttt{<e>} & \texttt{[ans]} \\
  C ($N{=}5$) & \texttt{<s>} & \texttt{<t>} & \texttt{<t>} & \texttt{<t>} & \texttt{<t>} & \texttt{<t>} & \texttt{<e>} & \texttt{[ans]} \\
  \midrule
  Attn (A) & $1$ & $1$ & $1$ & $0$ & $0$ & $0$ & $1$ & $1$ \\
  Attn (B) & $1$ & $1$ & $1$ & $1$ & $0$ & $0$ & $1$ & $1$ \\
  Attn (C) & $1$ & $1$ & $1$ & $1$ & $1$ & $1$ & $1$ & $1$ \\
  \bottomrule
  \end{tabular}%
  }
  \caption{Batch layout for variable $N$.}
  \label{fig:batch_layout}
\vspace{-1em}

  \end{figure}
\textbf{Batch-Efficient Variable Depth Processing.}
Adaptive $N$ naturally produces per-sample reasoning depths within a batch. We design a padding-and-masking scheme that unifies all samples into a single $\max(N)$-iteration latent loop: short-$N$ samples receive masked pad tokens whose attention weights are zeroed, so the loop runs identically for all samples without branching, retaining full GPU parallelism. Figure~\ref{fig:batch_layout} illustrates this layout, with further details in Appendix~\ref{sec:app_batch}.

\subsection{SFT Phase: Building Semantically Anchored Latent Reasoning (Challenge 1 \& 2)}
\label{sec:sft}

The SFT phase addresses the first two challenges identified in the Introduction: (1) the semantic grounding gap between prior-less SIDs and latent reasoning, and (2) representation drift. With the latent reasoning mechanism defined above, we now describe how it is trained.

\begin{figure}[t]
\centering
\includegraphics[width=0.65\linewidth]{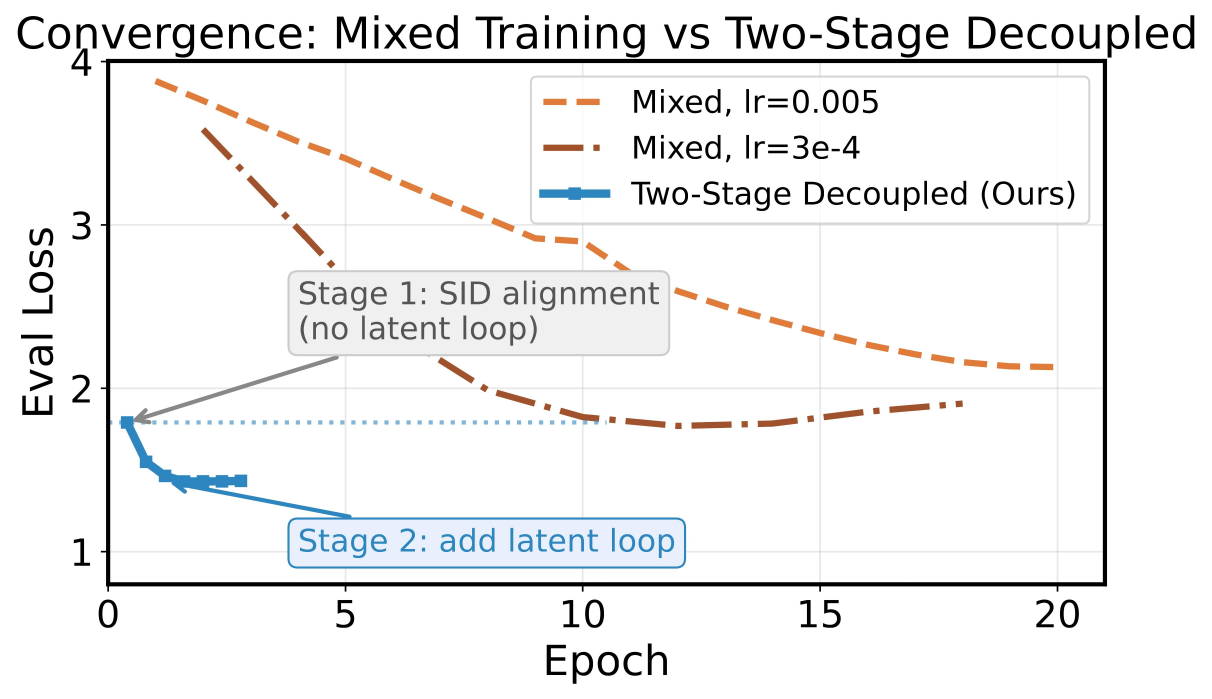}
\vspace{-1em}
\caption{Two-stage decoupling vs.\ mixed training.}
\label{fig:convergence}
\vspace{-2em}

\end{figure}
\textbf{Why decoupling is necessary.} Unlike NLP tokens that carry pre-trained semantic priors, SID tokens are constructed from scratch with zero prior semantics. Joint training forces the model to simultaneously ground a new symbolic system \emph{and} perform continuous-space reasoning, and without semantic anchors latent reasoning has no stable trajectory to follow. Figure~\ref{fig:convergence} confirms this: mixed training starts with evaluation loss as high as 3.5--3.9 and converges extremely slowly: after 10 epochs, loss remains above 1.8 (lr=$3{\times}10^{-4}$) or 2.9 (lr=$5{\times}10^{-3}$). Counterintuitively, increasing the learning rate from $3{\times}10^{-4}$ to $5{\times}10^{-3}$ slows convergence rather than accelerating it, suggesting that the two objectives actively interfere rather than merely lacking optimization capacity.

We therefore adopt a \textbf{two-stage decoupling} strategy. \textbf{Stage~1} serves as a semantic grounding phase: the model learns to generate the recommended item's SID via cross-entropy loss, establishing the SID-to-item semantic mapping so that each symbol acquires stable meaning. After Stage~1 converges, \textbf{Stage~2} introduces the latent reasoning mechanism (Section~\ref{sec:latent_mechanism}) with CoT semantic alignment, now able to reason on a stable semantic foundation. In contrast to mixed training, two-stage decoupling starts from 1.79 (Stage~1 has already grounded SID semantics), and Stage~2 converges to $\approx$1.44 within 4 epochs, reducing total training time from 20+ hours to approximately 6 hours.

While decoupling resolves the semantic grounding gap (Challenge~1), representation drift (Challenge~2) remains: latent reasoning in continuous space lacks the natural constraints of discrete tokens, and hidden states can drift toward meaningless representations without explicit semantic guidance. We address this through CoT semantic alignment, anchoring the latent reasoning trajectory to explicit CoT segments.

\textbf{CoT Semantic Alignment.}\label{sec:sft_align} Ablation (Table~\ref{tab:sft_ablation}) confirms representation drift: naively adding latent reasoning without alignment degrades performance. To prevent this, we anchor each latent reasoning step to explicit CoT semantic segments during SFT, and switch to terminal-only alignment during RL (Section~\ref{sec:rl}) to support variable-length reasoning.

\textbf{Explicit CoT Anchor Construction.} The key insight is that explicit CoT reasoning provides a natural trajectory for latent reasoning to follow. Concretely, a large model (e.g., GPT) generates CoT reasoning text per training sample for alignment supervision only. At inference, LASAR requires no CoT text. The CoT text is semantically segmented using an embedding model (e.g., bge-small-en-v1.5)~\cite{64_bge_embedding}. The same backbone model then encodes each segment (offline, before training) and extracts the last-token hidden state from the final Transformer layer as pre-computed alignment anchors. This design shares similarities with CODI~\cite{60_CODI}'s self-distillation, as both guide latent reasoning through hidden states from explicit reasoning, but with a key difference: CODI uses L1 loss for single-token alignment only at the answer generation position, whereas we perform \textbf{multi-step alignment} between each latent step and the corresponding explicit CoT segment during SFT, using bidirectional KL divergence between their softmax-normalized activation profiles. The number of resulting segments also serves as the supervision label for the Policy Head's step prediction.

\textbf{Step-wise Bidirectional KL Alignment.} The SFT phase adopts a step-wise alignment strategy: aligning each latent step's hidden state with the corresponding explicit CoT segment's hidden state via bidirectional KL divergence:
{\small
\vspace{-4pt}
\begin{equation}
L_{\text{align}} = \frac{1}{N}\sum_{t=1}^{N} D_{\text{KL}}^{\text{bidir}}(h_t, h_t^{\text{cot}})
\end{equation}
\vspace{-4pt}

}
where $D_{\text{KL}}^{\text{bidir}}(a, b) = \frac{1}{2}\big(D_{\text{KL}}(\text{Softmax}(a) \| \text{Softmax}(b)) + D_{\text{KL}}(\text{Softmax}(b) \| \text{Softmax}(a))\big)$, $h_t$ is the hidden state from latent step $t$ (Eq.~\ref{eq:latent_loop}), and $h_t^{\text{cot}}$ is the hidden state obtained by encoding the $t$-th CoT segment through the same backbone.

Bidirectional KL compares the relative activation profiles of latent and CoT states in the shared backbone space. Ablation (Section~\ref{sec:ablation}) confirms it is the only alignment method among the compared alternatives yielding positive improvement.

\textbf{SFT Total Loss.} The Policy Head is trained with cross-entropy loss using the number of CoT semantic segments as labels for warm start. The total SFT loss combines all objectives: $L_{\text{SFT}} = L_{\text{CE}} + \alpha_{\text{align}} \cdot L_{\text{align}} + \beta_{\text{policy}} \cdot L_{\text{policy}}$, where $L_{\text{CE}}$ is the SID generation loss (answer tokens only), $L_{\text{policy}}$ is the Policy Head CE loss, and $L_{\text{align}}$ is the step-wise bidirectional KL loss.
\subsection{RL Phase: Joint Quality and Efficiency Optimization (Challenge 3)}
\label{sec:rl}

The SFT phase provides the Policy Head with initial step prediction via CE loss, but fixed CoT segment labels do not directly optimize recommendation quality or reasoning efficiency. The RL phase addresses this through three coordinated objectives: GRPO (generation quality), REINFORCE (step optimization), and Terminal KL (semantic alignment).

\textbf{GRPO for Recommendation Quality.}
For each prompt, $G$ candidates are generated, and the reward combines exact match and ranking quality following prior work~\cite{37_MiniOneRec}: $r = r_{\text{rule}} + r_{\text{NDCG}}$, where $r_{\text{rule}}$ is binary (1 if hit, 0 otherwise) and $r_{\text{NDCG}}$ penalizes non-target candidates by ranking position (formal definitions in Appendix~\ref{sec:app_reward}). The clipped GRPO objective with KL penalty is:
{\small
\begin{align}
L_{\text{GRPO}} = {} & -\mathbb{E}\Big[\min\big(\rho_i(\Theta)\,\hat{A}_i,\; \text{clip}(\rho_i(\Theta), 1{-}\varepsilon, 1{+}\varepsilon)\,\hat{A}_i\big)\Big] \nonumber \\
& + \beta \, D_{\text{KL}}\big(\pi_\Theta \,\|\, \pi_{\text{ref}}\big)
\end{align}
}
where $\pi_\Theta$ is the backbone policy, $\pi_{\text{ref}}$ a frozen reference (KL only), $\pi_{\Theta_{\text{old}}}$ the rollout policy, $\rho_i(\Theta) = \frac{\pi_\Theta(y_i|x)}{\pi_{\Theta_{\text{old}}}(y_i|x)}$, $\varepsilon$ is the clipping ratio, and $\hat{A}_i = \frac{r_i - \text{mean}(\{r_j\}_{j=1}^G)}{\text{std}(\{r_j\}_{j=1}^G)+\epsilon}$ is the group-normalized advantage~\cite{53_DeepSeekMath} ($\epsilon{=}10^{-4}$ for numerical stability; see Appendix~\ref{sec:app_reward}).

\textbf{REINFORCE for Adaptive Step Allocation.}
\label{sec:reinforce}
The CE-trained Policy Head provides a warm-start step distribution, but the label distribution is determined by the segmenter's granularity rather than by what benefits recommendation quality. REINFORCE directly optimizes step allocation through reward-guided exploration, enabling the Policy Head to learn more efficient distributions that improve recommendation quality while reducing average step count. During the RL phase, the Policy Head samples $N \sim \pi_\phi(\cdot \mid h_0)$, and the sampling strategy is optimized via the REINFORCE algorithm:
{\small
\begin{align}
L_{\text{REINFORCE}} = {} & -\mathbb{E}_{N \sim \pi_\phi}\!\left[(R_{\text{group}} - b_{\text{EMA}} - \lambda N)\, \log \pi_\phi(N \mid h_0)\right] \nonumber \\
& - \eta \cdot H(\pi_\phi)
\end{align}
}
where $R_{\text{group}}$ is the group-level reward for the current prompt (averaged over $G$ candidates), $b_{\text{EMA}}$ is the exponential moving average baseline of reward (reducing variance), $\lambda N$ is the step count penalty (encouraging efficiency), and $H(\pi_\phi)$ is an entropy regularization term with coefficient $\eta$ (preventing degeneration to a less diverse step count). Switching from argmax to sampling enables REINFORCE to learn better step allocation through exploration. The warm-start distribution established by SFT ensures meaningful exploration starting points rather than starting from a random policy.

\textbf{Terminal Alignment for Variable-Length Reasoning.}
The SFT phase's step-wise alignment requires each step to correspond to a fixed CoT segment, which is no longer applicable during RL: $N$ is dynamically sampled, producing variable-length reasoning chains that cannot be aligned step-by-step to the fixed-length CoT anchors.

Therefore, the RL phase switches to a \textbf{Terminal-only strategy}: aligning only at the \textbf{last} latent step of the reasoning chain, $L_{\text{Terminal KL}} = D_{\text{KL}}^{\text{bidir}}(h_N, h_{\text{final}}^{\text{cot}})$, where $h_N$ is the final latent hidden state from Eq.~\ref{eq:latent_loop} and $h_{\text{final}}^{\text{cot}}$ is the hidden state of the last explicit CoT segment, encouraging the reasoning endpoint to remain semantically aligned with the final CoT anchor regardless of $N$.

\textbf{RL Total Loss.} The three components are combined as $L_{\text{total}} = L_{\text{GRPO}} + \gamma_{\text{KL}} \cdot L_{\text{Terminal KL}} + \gamma_{\text{RF}} \cdot L_{\text{REINFORCE}}$.
\section{Experiments}
\label{sec:experiments}
We evaluate LASAR on three Amazon product review datasets to answer four questions: \textbf{(RQ1)} Does LASAR outperform traditional, generative, latent-reasoning, and explicit-CoT methods (Section~\ref{sec:main_results})? \textbf{(RQ2)} What are the individual contributions of latent reasoning, alignment, and REINFORCE (Section~\ref{sec:ablation})? \textbf{(RQ3)} How does adaptive step allocation compare with fixed reasoning depths (Section~\ref{sec:step_opt})? \textbf{(RQ4)} What are LASAR's efficiency and scaling properties (Section~\ref{sec:efficiency})?

\begin{table*}[t]
  \centering
  \caption{Main results across three Amazon datasets. Best in \textbf{bold}, second-best \underline{underlined}.}
  \label{tab:main_results}
  \renewcommand{\arraystretch}{0.85}
  \small
  \begin{tabular*}{0.95\textwidth}{@{\extracolsep{\fill}}llcccccc@{}}
  \toprule
  & Model & N@5 & N@10 & N@20 & HR@5 & HR@10 & HR@20 \\
  \midrule
  \multirow{7}{*}{\rotatebox[origin=c]{90}{Sports}}
  & \textbf{LASAR}     & \textbf{0.0121} & \textbf{0.0152} & \textbf{0.0188} & \textbf{0.0185} & \textbf{0.0280} & \textbf{0.0425} \\
  & Explicit CoT$_{\text{GREAM}}$                  & 0.0089 & 0.0118 & \underline{0.0153} & 0.0138 & 0.0228 & \underline{0.0370} \\
  & MiniOneRec                  & \underline{0.0099} & \underline{0.0126} & 0.0152 & \underline{0.0155} & \underline{0.0237} & 0.0339 \\

  & ReaRec                       & 0.0086 & 0.0112 & 0.0143 & 0.0151 & 0.0233 & 0.0355 \\
  & LC-Rec                           & 0.0081 & 0.0100 & 0.0118 & 0.0123 & 0.0184 & 0.0254 \\
  & GRU4Rec                          & 0.0062 & 0.0080 & 0.0101 & 0.0090 & 0.0147 & 0.0232 \\
  & SASRec                           & 0.0060 & 0.0074 & 0.0094 & 0.0089 & 0.0132 & 0.0212 \\
  \midrule
  \multirow{7}{*}{\rotatebox[origin=c]{90}{Instruments}}
  & \textbf{LASAR}                 & \textbf{0.0612} & \textbf{0.0667} & \textbf{0.0730} & \textbf{0.0763} & \textbf{0.0937} & \textbf{0.1184} \\
  & Explicit CoT$_{\text{GREAM}}$                  & 0.0574 & 0.0621 & 0.0674 & 0.0703 & 0.0850 & 0.1060 \\
  & MiniOneRec                       & \underline{0.0604} & \underline{0.0640} & \underline{0.0677} & \underline{0.0715} & 0.0826 & 0.0974 \\

  & ReaRec                       & 0.0494 & 0.0548 & 0.0604 & 0.0705 & \underline{0.0873} & \underline{0.1095} \\
  & LC-Rec                           & 0.0533 & 0.0561 & 0.0587 & 0.0616 & 0.0701 & 0.0803 \\
  & SASRec                           & 0.0449 & 0.0475 & 0.0502 & 0.0536 & 0.0617 & 0.0725 \\
  & GRU4Rec                          & 0.0422 & 0.0454 & 0.0489 & 0.0527 & 0.0629 & 0.0769 \\
  \midrule
  \multirow{7}{*}{\rotatebox[origin=c]{90}{Beauty}}
  & \textbf{LASAR}         & \textbf{0.0239} & \textbf{0.0303} & \textbf{0.0366} & \textbf{0.0365} & \textbf{0.0563} & \underline{0.0813} \\
  & Explicit CoT$_{\text{GREAM}}$                  & 0.0228 & 0.0293 & \underline{0.0365} & 0.0351 & \underline{0.0553} & \textbf{0.0837} \\

  & MiniOneRec                       & \underline{0.0232} & \underline{0.0295} & 0.0358 & \underline{0.0352} & 0.0542 & 0.0795 \\

  & ReaRec                       & 0.0201 & 0.0255 & 0.0307 & 0.0296 & 0.0464 & 0.0673 \\

  & LC-Rec                           & 0.0178 & 0.0222 & 0.0261 & 0.0260 & 0.0407 & 0.0592 \\
  & SASRec                           & 0.0159 & 0.0195 & 0.0229 & 0.0232 & 0.0343 & 0.0480 \\
  & GRU4Rec                          & 0.0144 & 0.0190 & 0.0242 & 0.0226 & 0.0370 & 0.0573 \\
  \bottomrule
  \end{tabular*}%
  \vspace{-10pt}
  \end{table*}
\subsection{Experimental Setup}
\textbf{Datasets.} We evaluate on three Amazon product review datasets~\cite{63_AmazonReviews2018}: Beauty, Instruments, and Sports. We apply 5-core filtering and leave-one-out evaluation, following prior works~\cite{28_SASRec,29_BERT4Rec,30_GRU4Rec,34_TIGER,36_LCRec,37_MiniOneRec}. Sparsity ranges from 99.935\% to 99.984\%; detailed statistics are in Table~\ref{tab:dataset} (Appendix~\ref{sec:app_dataset}).

\textbf{Baselines.} We compare against six baselines spanning four categories: traditional sequential models (SASRec~\cite{28_SASRec}, GRU4Rec~\cite{30_GRU4Rec}), LLM-based generative methods (LC-Rec~\cite{36_LCRec}, MiniOneRec~\cite{37_MiniOneRec}), latent reasoning (ReaRec~\cite{46_ReaRec}), and explicit CoT reasoning (GREAM~\cite{41_GREAM}). All generative baselines share the same base model, prompt, and training data; performance differences thus reflect the reasoning mechanism alone. For GREAM, we retain its core CoT chain but strip the 19 augmentation prompts to MiniOneRec's prompt and adopt its own ablation's strongest configuration denoted Explicit CoT$_{\text{GREAM}}$. See Appendix~\ref{sec:app_baselines} for details.

\textbf{Evaluation Metrics.} We report NDCG@$K$ and Hit Rate@$K$ for $K \in \{5, 10, 20\}$ (Appendix~\ref{sec:app_metrics}). Beam search width is set to 50 for all generative methods; at inference, the model autoregressively generates $M$ SID tokens per candidate, and beam search ensures that the top-$K$ items are ranked by joint token probability.

\textbf{Implementation Details.} All generative methods share the same SID encoding, training data, and prompt format. CoT reasoning text follows GREAM's structured format, while the input prompt is simplified to MiniOneRec's template. We use Qwen3-0.6B~\cite{qwen3} as the base model and further scale to 1.7B (Section~\ref{sec:scaling}), optimized with AdamW (cosine LR schedule with 0.08 warmup). Each item is represented as $M{=}4$ SID tokens from 256-entry codebooks (up to $256^4$ sequences). Teacher CoT reasoning text is generated by GPT-5 (see Appendix~\ref{sec:app_prompt_cot}). All experiments run on $8{\times}$L40 (48\,GB); full hyperparameters are in Table~\ref{tab:implementation} (Appendix~\ref{sec:app_engineering}).
\subsection{Main Results (RQ1)}
\label{sec:main_results}


Table~\ref{tab:main_results} presents the results across all three datasets under NDCG and Hit Rate at $K{=}5,10,20$. Among all methods, LASAR achieves the best performance on nearly all metric--dataset combinations, with the sole exception of Beauty HR@20 where Explicit CoT yields marginal gains at high recall cutoffs. On Sports (most sparse), LASAR achieves particularly strong gains; on Instruments, it shows consistent improvements that widen at larger $K$ ($+7.8\%$ NDCG@20 over MiniOneRec), and on Beauty gains reach $+3.9\%$ HR@10 over MiniOneRec, suggesting latent reasoning is especially beneficial under high sparsity, where the model's semantic understanding compensates for limited collaborative signals. Explicit CoT's limited gains may stem from representation interference between language modeling and collaborative filtering objectives, while LASAR reduces such competition by keeping intermediate reasoning in continuous latent space. Bootstrap tests show significance on Sports and Instruments ($p < 0.05$), and marginal significance on Beauty ($p < 0.1$, except $K{=}20$).
\subsection{Ablation Studies (RQ2)}
\label{sec:ablation}

We next decompose LASAR to understand which components drive its gains (Tables~\ref{tab:sft_ablation}, \ref{tab:ablation}).

\textbf{SFT Phase Ablation.} Table~\ref{tab:sft_ablation} compares different alignment strategies during the SFT phase on Beauty (see Table~\ref{tab:sft_cross} for Sports and Instruments). Latent reasoning without alignment yields almost no improvement over the Pure SFT baseline ($+$0.4\% NDCG@10), confirming that \textbf{latent reasoning requires alignment to be effective}. Among the three alignment variants, only KL divergence yields positive improvement, whereas Cosine and MSE both underperform even the \emph{no-alignment} baseline. Unlike cosine similarity or MSE, bidirectional KL directly aligns the softmax-normalized relative activation profiles.

\begin{table*}[htbp]
  \centering
  \caption{SFT-phase ablation on Beauty: alignment methods.}
  \vspace{-3pt}
  \label{tab:sft_ablation}
  \resizebox{0.8\textwidth}{!}{%
  \begin{tabular}{lcccccccr}
  \toprule
  Model & Two-Stage & Latent & Alignment & N@5 & N@10 & HR@5 & HR@10 & $\Delta$ N@10 \\
  \midrule
  Pure SFT (MiniOneRec) & & & -- & 0.0212 & 0.0277 & 0.0329 & 0.0531 & -- \\
  + Latent (w/o align.) & $\checkmark$ & $\checkmark$ & None & 0.0207 & 0.0278 & 0.0327 & 0.0550 & $+$0.4\% \\
  \textbf{+ KL Alignment} & $\checkmark$ & $\checkmark$ & \textbf{KL} & \textbf{0.0217} & \textbf{0.0285} & \textbf{0.0340} & \textbf{0.0552} & \textbf{+2.9\%} \\
  + Cosine Alignment & $\checkmark$ & $\checkmark$ & Cosine & 0.0211 & 0.0277 & 0.0341 & 0.0543 & 0.0\% \\
  + MSE Alignment & $\checkmark$ & $\checkmark$ & MSE & 0.0187 & 0.0245 & 0.0295 & 0.0477 & $-$11.6\% \\
  \bottomrule
  \end{tabular}%
  }
  \vspace{-4pt}

  \end{table*}

\textbf{RL-Phase Ablation.} The SFT ablation confirms that KL alignment enables latent reasoning to work, and we now assess the distinct contribution of each of the three RL components (GRPO, Terminal KL, REINFORCE). Table~\ref{tab:ablation} presents the RL-phase ablation on Beauty, chosen for its large interaction count (176K) that provides stable gradient signals. Key findings are further validated on Sports in Section~\ref{sec:step_opt}.

\begin{table*}[t]
  \centering
  \caption{RL-phase ablation on Beauty. $\Delta$: relative to previous row.}
  \vspace{-4pt}
  \label{tab:ablation}
  \resizebox{0.9\textwidth}{!}{%
  \begin{tabular}{lccccccccr}
  \toprule
  Model & Latent & Terminal KL & REINFORCE & N@5 & N@10 & HR@5 & HR@10 & Mean $N$ & $\Delta$ N@10 \\
  \midrule
  MiniOneRec & N/A & N/A & N/A & 0.0232 & 0.0295 & 0.0352 & 0.0542 & N/A & -- \\
  RL w/ latent reasoning & $\checkmark$ & & & 0.0227 & 0.0287 & 0.0346 & 0.0533 & 3.59 & $-$2.7\% \\
  + Terminal KL Alignment & $\checkmark$ & $\checkmark$ & & 0.0233 & 0.0294 & 0.0353 & 0.0543 & 4.20 & +2.4\% \\
  \textbf{+ REINFORCE (LASAR)} & $\checkmark$ & $\checkmark$ & $\checkmark$ & \textbf{0.0239} & \textbf{0.0303} & \textbf{0.0365} & \textbf{0.0563} & \textbf{2.47} & \textbf{+3.1\%} \\
  \bottomrule
  \end{tabular}%
  }
  \vspace{-6pt}
  \end{table*}

Naively adding latent reasoning without proper RL training degrades performance, confirming \textbf{latent reasoning is not a free improvement}. Terminal KL alignment recovers this drift and improves NDCG@10, but also increases Mean $N$: alignment makes each step productive, and without REINFORCE's step penalty there is no incentive to compress depth. Adding REINFORCE further improves NDCG@10 while compressing Mean $N$ from $\approx$4.2 to $\approx$2.5, demonstrating \textbf{step compression and quality improvement are concurrent}. Terminal KL and REINFORCE each provide independent and complementary gains.
\subsection{Step Optimization Analysis (RQ3)}
\label{sec:step_opt}

The ablation showed that REINFORCE effectively compresses reasoning steps while improving quality. We now investigate this mechanism in detail. We note that the SFT supervision labels (Teacher CoT segment counts) correlate positively with sample complexity: on Sports, 83.4\% of samples have 3 segments (avg.\ history length 6.9, category diversity 6.2), 16.3\% have 4 segments (hist.\ 8.5, cat.\ div.\ 7.5), and 0.2\% have 5 segments (hist.\ 9.8, cat.\ div.\ 8.8) (Table~\ref{tab:teacher_cot} in Appendix~\ref{sec:app_sft_labels}), confirming the Policy Head learns meaningful difficulty-aware depth allocation during SFT.

\begin{figure}[h]
\centering
\vspace{-5pt}
\begin{subfigure}[b]{0.49\linewidth}
  \centering
  \includegraphics[width=\linewidth]{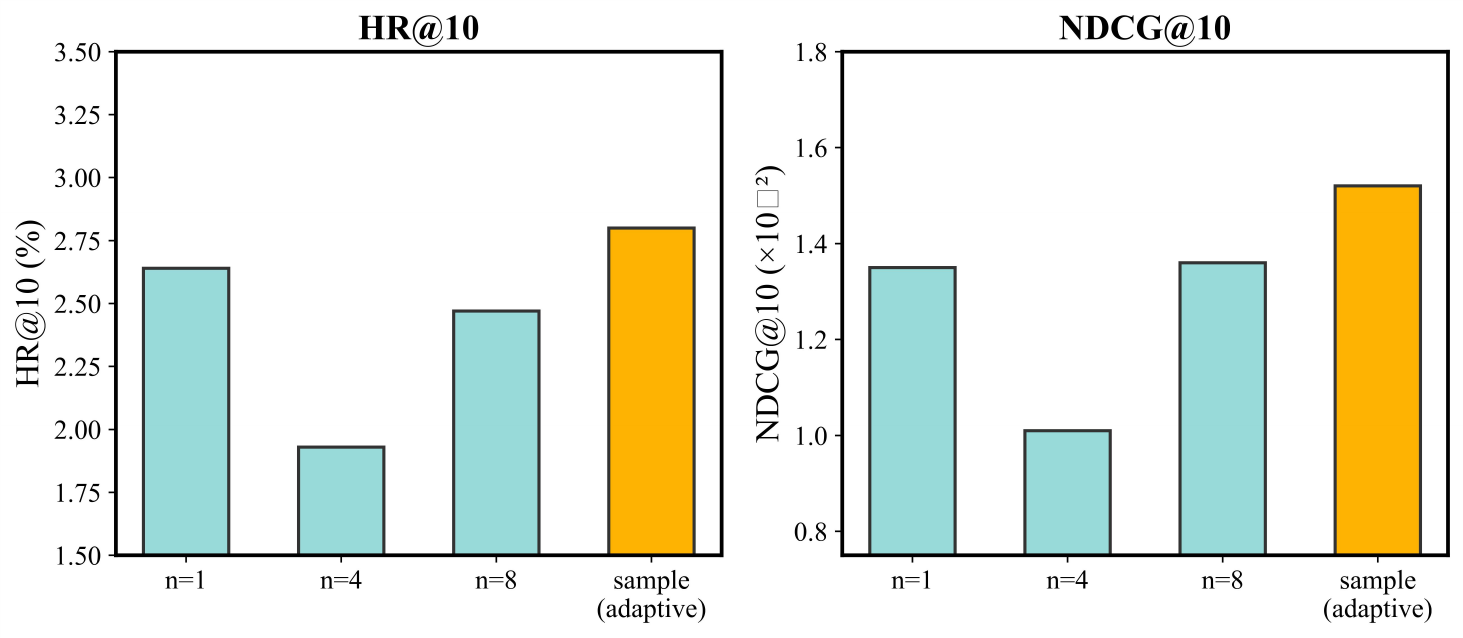}
  \caption{Force $N$ comparison.}
  \label{fig:force_n}
\end{subfigure}
\hfill
\begin{subfigure}[b]{0.49\linewidth}
  \centering
  \includegraphics[width=\linewidth]{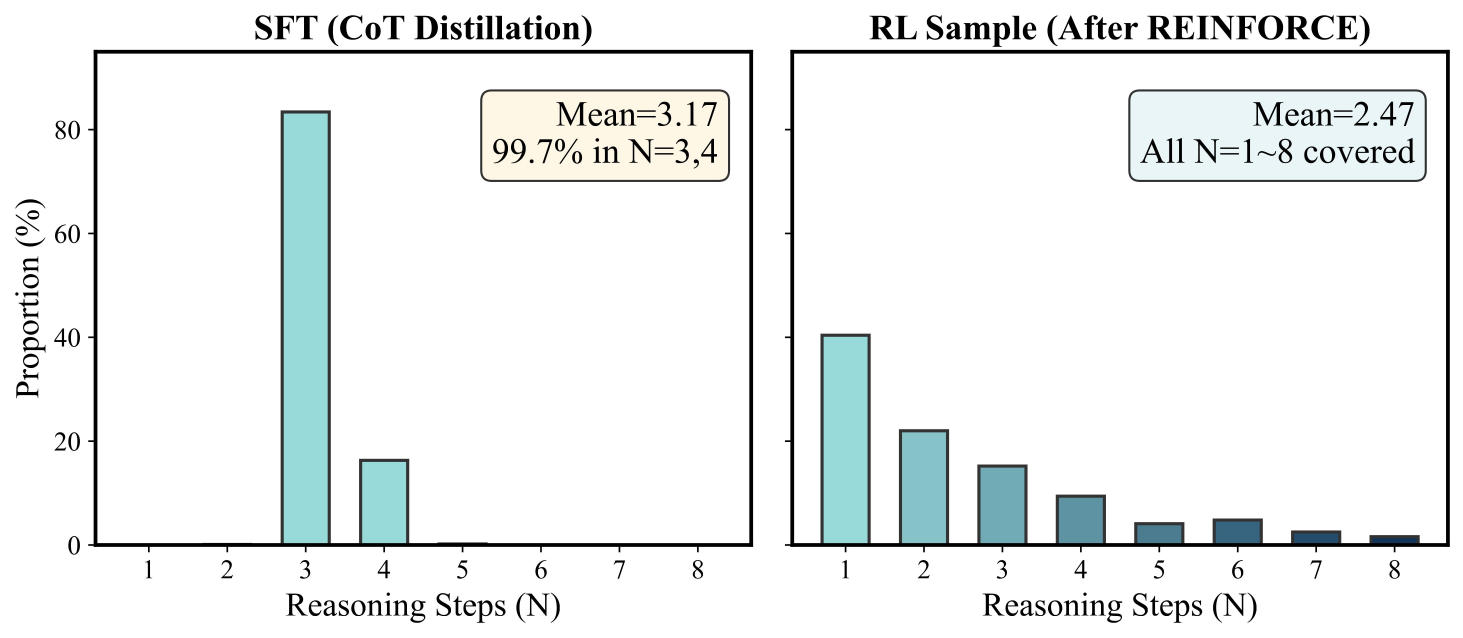}
  \caption{Step distribution: SFT vs.\ RL.}
  \label{fig:distribution}
\end{subfigure}
\vspace{-5pt}
\caption{Adaptive step allocation on Sports: adaptive $N$ outperforms all fixed reasoning depths, and RL redistributes steps toward fewer but better-allocated depths.}
\label{fig:force_and_distribution}
\vspace{-13pt}
\end{figure}

\textbf{Force $N$ Experiment: Adaptive Allocation Outperforms Fixed Depths.}
Figure~\ref{fig:force_n} compares adaptive sampling against three fixed-$N$ configurations on Sports. Adaptive (HR@10 $= 2.80\%$) surpasses all fixed reasoning depths. Fixed $N{=}4$ performs worst (1.93\%): forcing all samples through 4 latent iterations introduces interference, whereas $N{=}1$ preserves representations without disruption.

\begin{figure}[h]
  \centering
  \vspace{-4pt}
  \begin{subfigure}[b]{0.49\linewidth}
    \centering
    \includegraphics[width=\linewidth]{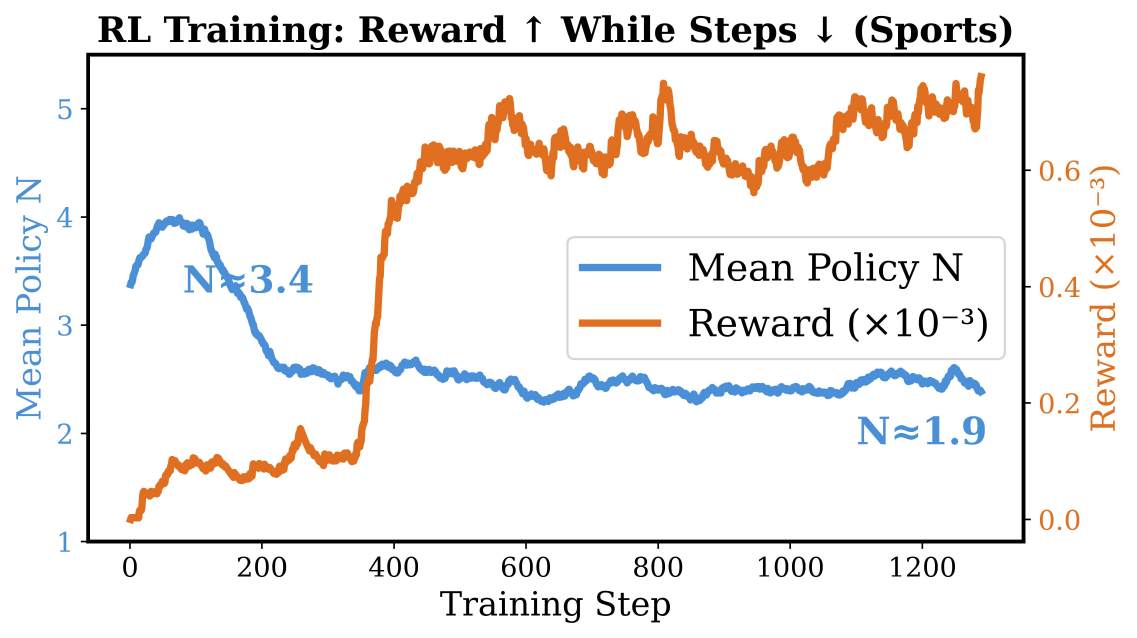}
    \caption{RL training dynamics.}
    \label{fig:rl_dynamics}
  \end{subfigure}
  \hfill
  \begin{subfigure}[b]{0.49\linewidth}
    \centering
    \includegraphics[width=\linewidth]{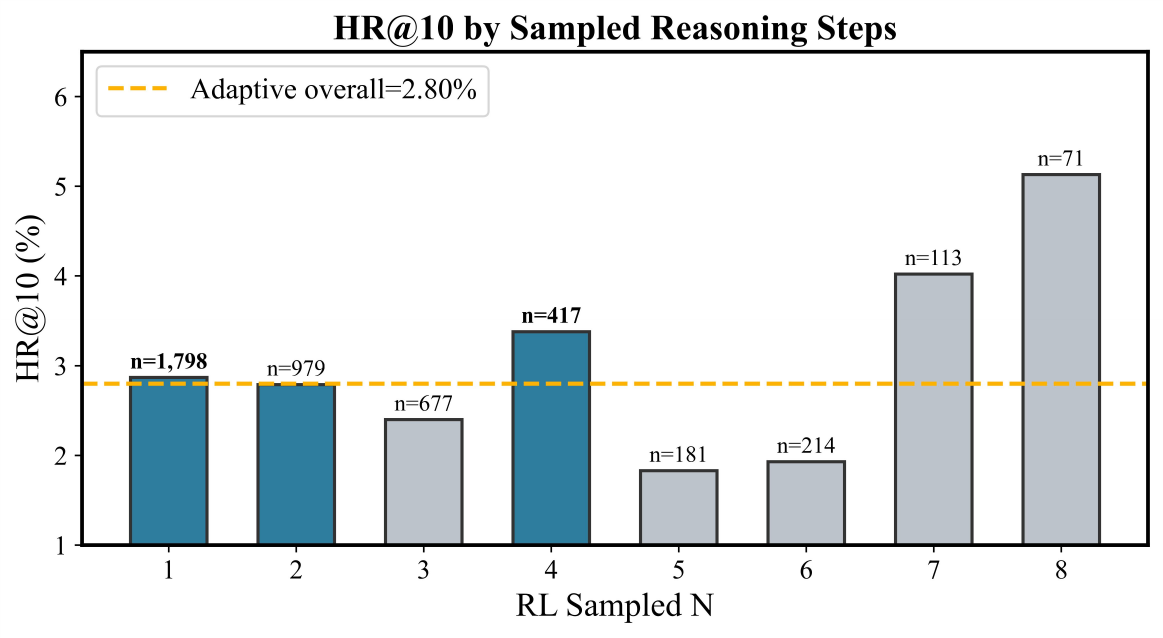}
    \caption{Per-$N$ HR@10.}
    \label{fig:per_n}
  \end{subfigure}
  \vspace{-5pt}
  \caption{(a) RL training dynamics on Sports: Mean $N$ drops early then stabilizes while Reward rises. (b) Per-$N$ HR@10: non-monotonic relationship with a peak at $N{=}4$ among mid-range depths, as the Policy Head selectively assigns deeper reasoning to samples that benefit.}
  \label{fig:rl_dynamics_and_per_n}
  \vspace{-8pt}
  \end{figure}

\textbf{RL Dynamics and Per-$N$ Analysis.}
Figure~\ref{fig:rl_dynamics} tracks the RL training dynamics on Sports: Mean $N$ drops sharply from ${\approx}3.4$ to ${\approx}1.9$ in early training and stabilizes, while Reward rises consistently. REINFORCE thus compresses the average step budget without sacrificing quality (indeed improving it) (Table~\ref{tab:ablation}). Figure~\ref{fig:distribution} shows the distribution shifts from SFT's concentrated $N{=}3,4$ (99.7\%) to a broader $N{=}1$--8 coverage (Mean $= 2.47$).

REINFORCE does not simply minimize depth. Per-$N$ analysis (metrics per step count; Figure~\ref{fig:per_n}) reveals that the Policy Head learns a selective allocation strategy: most samples receive shallow reasoning ($N{\leq}4$) for efficiency, while the hardest cases are assigned deep reasoning ($N{\geq}7$). The intermediate depths ($N{=}5,6$) see few samples and lower HR@10, as they are neither efficient nor thorough enough. The key evidence is the gap between Force~$N$ and per-$N$ results: forcing all samples to $N{=}4$ yields the worst result (1.93\%), yet the Policy Head achieves 3.38\% at $N{=}4$ for selected samples, confirming it identifies which samples genuinely benefit from deeper reasoning.

\vspace{-4pt}
\subsection{Inference Efficiency and Model Scaling (RQ4)}
\label{sec:efficiency}
Beyond recommendation quality, latent reasoning avoids the latency overhead of explicit text generation. We evaluate LASAR's inference efficiency and scalability to larger models.
\begin{table}[h]
  \centering
  \vspace{-10pt}
  \caption{Inference efficiency.}
  \label{tab:efficiency}
  \resizebox{\linewidth}{!}{%
  \begin{tabular}{llrr}
  \toprule
  Dataset & Method & GPU-sec/Sample & Wall-clock (8$\times$L40) \\
  \midrule
  \multirow{3}{*}{Beauty}
  & MiniOneRec & 0.27 & 12min \\
  & \textbf{LASAR} & 0.29 & 13min \\
  & Explicit CoT$_{\text{GREAM}}$ (CoT Gen.) & 7.0 & 5.5h \\
  \midrule
  \multirow{3}{*}{Instruments}
  & MiniOneRec & 0.25 & 13min \\
  & \textbf{LASAR} & 0.29 & 15min \\
  & Explicit CoT$_{\text{GREAM}}$ (CoT Gen.) & 6.5 & 5h \\
  \midrule
  \multirow{3}{*}{Sports}
  & MiniOneRec & 0.30 & 22min \\
  & \textbf{LASAR} & 0.32 & 24min \\
  & Explicit CoT$_{\text{GREAM}}$ (CoT Gen.) & 7.0 & 8.5h \\
  \bottomrule
  \end{tabular}%
  }
  \vspace{-5pt}
  \end{table}
\vspace{-4pt}

\textbf{Inference Efficiency.} Table~\ref{tab:efficiency} summarizes inference latency across all three datasets under beam width 50 on 8$\times$L40. LASAR adds only modest inference overhead over MiniOneRec ($\approx$7--16\%). Explicit CoT$_{\text{GREAM}}$ has two inference modes: our main results (Table~\ref{tab:main_results}) use its stronger \emph{direct-answer} mode (no CoT text; latency comparable to MiniOneRec), while Table~\ref{tab:efficiency} reports its \emph{CoT-generation} mode for when interpretable traces are required. LASAR is more accurate than direct-answer GREAM and over 20$\times$ faster than its CoT-generation mode, whose overhead comes from autoregressively decoding long reasoning chains, yielding a favorable efficiency--effectiveness trade-off.

\begin{table}[h]
  \centering
  \caption{Model scaling on Beauty.}
  \vspace{-6pt}
  \label{tab:scale_up}
  \resizebox{0.8\linewidth}{!}{%
  \begin{tabular}{lcccc}
  \toprule
  & \multicolumn{2}{c}{0.6B Full FT} & \multicolumn{2}{c}{1.7B LoRA} \\
  \cmidrule(lr){2-3} \cmidrule(lr){4-5}
  Method & N@10 & HR@10 & N@10 & HR@10 \\
  \midrule
  \textbf{LASAR} & \textbf{0.0303} & \textbf{0.0563} & \textbf{0.0307} & \textbf{0.0592} \\
  MiniOneRec & 0.0295 & 0.0542 & 0.0299 & 0.0556 \\
  Explicit CoT$_{\text{GREAM}}$ & 0.0293 & 0.0553 & 0.0295 & 0.0561 \\
  \bottomrule
  \end{tabular}%
  }
  \vspace{-3pt}
  \end{table}
  
\textbf{Model Scaling.} \label{sec:scaling}
To assess scalability, we scale from the default Qwen3-0.6B to Qwen3-1.7B (LoRA) on Beauty, both trained with SFT+RL. Table~\ref{tab:scale_up} shows LASAR achieves the best performance at both scales. Both LASAR and MiniOneRec improve comparably in N@10, confirming that latent reasoning does not limit capacity gains. LASAR's HR@10 gain from 0.6B to 1.7B also exceeds that of both baselines.
\vspace{-6pt}
\section{Conclusion}
\label{sec:conclusion}
This paper proposes LASAR, the first framework to perform complete latent reasoning with recurrent hidden-state feedback and per-sample adaptive depth in mainstream generative recommendation. Through two-stage SFT decoupling, explicit CoT semantic alignment with bidirectional KL divergence, and REINFORCE-based adaptive step optimization, LASAR simultaneously improves recommendation quality and reduces inference cost. Experiments on three datasets demonstrate that LASAR achieves the best overall performance across the evaluated settings. Ablation confirms that semantic alignment, Terminal KL, and REINFORCE make distinct and complementary contributions.

\textbf{Limitations and future work.} A shared challenge of Coconut-style latent reasoning is that the hidden-state feedback loop precludes teacher forcing, requiring sequential, hard-to-parallelize forward passes. LASAR also depends on a teacher LLM for offline CoT synthesis, which never enters inference. Future work may target efficient latent-loop execution and extend to conversational recommendation.



\textbf{Generative AI Disclosure.} Generative AI tools assisted with language editing and drafting; the authors reviewed all outputs and retain full responsibility.

\bibliography{references}

\newpage
\appendix

\section{Related Work}
\label{sec:app_related}

\subsection{From Explicit to Latent LLM Reasoning}

LLM reasoning capability has become a central research direction in recent years. Wei et al.~\cite{01_CoT} first proposed Chain-of-Thought (CoT) prompting, significantly improving complex task performance by explicitly generating intermediate reasoning steps. Subsequently, Kojima et al.~\cite{02_ZeroShot} discovered that simply prompting ``Let's think step by step'' elicits zero-shot reasoning, while subsequent studies~\cite{03_SelfConsistency,04_ToT,05_GoT,06_CoTTheory,07_ExpressivePower} extended the boundaries of explicit CoT from angles including multi-sample majority voting, tree/graph-structured search, and theoretical expressiveness. However, the fundamental cost of explicit CoT lies in token-by-token reasoning text generation, with latency scaling linearly with chain length. Even as DeepSeek-R1~\cite{22_DeepSeekR1} and OpenAI O1~\cite{23_O1} have pushed explicit reasoning to its limits through RL, and Snell et al.~\cite{24_TestTimeCompute} provided optimal test-time compute allocation strategies, the generation cost of explicit reasoning remains a fundamental bottleneck.

To break this bottleneck, researchers have begun exploring the transfer of reasoning from the discrete token space to the continuous latent space. The core paradigm is: feed the hidden state from one LLM layer directly as the input embedding for the next step, iteratively refining reasoning through multi-step recurrent loops in continuous space, with intermediate states being unobservable dense vectors rather than readable discrete tokens. Under this definition, simply inserting additional learnable tokens or adding single-shot cross-attention does not constitute this form of latent reasoning.

Hao et al.~\cite{08_Coconut}'s Coconut is the pioneering work of this paradigm: feeding the last-layer hidden state as the next-step input embedding, realizing multi-step latent recurrent reasoning without generating any explicit tokens. Coconut adopts progressive curriculum training and discovers that latent space naturally supports BFS-like parallel path exploration. Its precursor works include Deng et al.~\cite{09_ImplicitCoT,10_FromExplicitCoTtoImplicitCoT}, which gradually internalize explicit reasoning into latent reasoning via knowledge distillation. Contemporaneously, Goyal et al.~\cite{11_PauseTokens} and Pfau et al.~\cite{12_ThinkDotByDot} showed that inserting learnable virtual tokens can allocate additional computation to the model, but such discrete-token reasoning is significantly weaker than Coconut's continuous recurrent representations~\cite{08_Coconut}. Cheng and Van Durme~\cite{13_CCoT} and Yu et al.~\cite{14_System2to1} also explored different paths of compressing explicit reasoning into latent space.

Along the Coconut line, follow-up work includes: Looped Transformers~\cite{15_LoopedTransformers} proving that the same set of parameters recurrently executed multiple times is equivalent to increasing model depth without adding parameters; Huginn~\cite{16_Huginn} adopting a prelude-core-coda architecture to realize deep recurrent latent reasoning, achieving performance comparable to larger models through test-time compute scaling; CODI~\cite{60_CODI} compressing explicit CoT into continuous space via self-distillation, matching CoT-SFT performance in a single training step. In diverse explorations, Heima~\cite{17_Heima} compresses each CoT step into a single thinking token; SoftCoT~\cite{18_SoftCoT} generates soft thinking tokens through a frozen LLM plus trainable projection module to avoid catastrophic forgetting; LaTRO~\cite{19_LaTRO} optimizes latent reasoning through variational inference and self-rewarding mechanisms. In adaptive reasoning, Think-at-Hard~\cite{20_ThinkAtHard} selectively triggers latent iterations based on token difficulty, and Token-Assorted~\cite{21_TokenAssorted} explores mixed reasoning with latent and text tokens.

In summary, latent reasoning has achieved significant progress in math and logic reasoning domains, but none of these methods have entered the mainstream decoder-only generative recommendation domain. This paper aims to fill exactly this gap.

\subsection{LLM-based Generative Recommendation}

Recommender systems are undergoing a shift from the traditional ID-embedding discriminative paradigm toward an LLM-driven generative paradigm. Traditional sequential recommendation methods such as SASRec~\cite{28_SASRec}, BERT4Rec~\cite{29_BERT4Rec}, and GRU4Rec~\cite{30_GRU4Rec} use self-attention or bidirectional encoders to extract preference representations from user interaction sequences, computing item scores via dot-product for ranking, a typical discriminative paradigm.

With the rise of LLMs, a series of works have reformulated recommendation as a sequence generation problem. Geng et al.'s P5~\cite{31_P5} and M6-Rec~\cite{32_M6Rec} pioneered a unified recommendation pretraining paradigm, unifying diverse recommendation signals into a text generation format. TALLRec~\cite{33_TALLRec} further explored parameter-efficient alignment of LLMs with recommendation. In the generative retrieval direction, Rajput et al.'s TIGER~\cite{34_TIGER} modeled recommendation as token-by-token generation of Semantic IDs, and subsequent work~\cite{35_IndexItemIDs} systematically studied semantic indexing methods for item IDs. Building on these foundations, LC-Rec~\cite{36_LCRec} integrated collaborative semantic information into LLMs, training them via SFT to directly generate recommended item IDs. Other studies~\cite{38_E4SRec,40_EAGER} advanced generative recommendation efficiency and effectiveness from perspectives of LLM-based sequential recommendation, single-LLM semantic tokenization, and behavior-semantic dual-stream collaboration, respectively. Recently, MiniOneRec~\cite{37_MiniOneRec} built the first fully open-source generative recommendation framework upon a streamlined OneRec, providing a complete LLM generative recommendation SFT+RL post-training pipeline and a solid foundation for our work.

In the reasoning-enhanced direction, GREAM~\cite{41_GREAM} proposed applying CoT explicit reasoning to generative recommendation, but the high generation latency of explicit CoT constrains its practical deployment efficiency. Other studies~\cite{42_RecSAVER,43_Reason4Rec,44_RecR1,45_R2EC,dai2025onepiecebringingcontextengineering} also explored combining LLM reasoning with recommender systems from different angles, but all rely on explicit textual reasoning chains. Overall, existing generative recommendation methods lack exploration of multi-step recurrent reasoning mechanisms in continuous latent space.

\subsection{Latent Reasoning in Recommendation}

Bringing the above latent reasoning paradigm into recommender systems is a frontier direction.

\textbf{Latent recurrent reasoning for traditional sequential recommendation.} ReaRec~\cite{46_ReaRec} realizes latent reasoning within a SASRec-style architecture, trained with temperature annealing and reasoning-aware contrastive learning, but uses globally fixed step counts. LARES~\cite{47_LARES} proposes a depth-recurrent latent reasoning framework, repeatedly refreshing all input token hidden states through the same set of Transformer layers at each step to increase computation density, with a two-stage strategy of self-supervised pretraining plus RL post-training. STREAMRec~\cite{48_STREAMRec} introduces stepwise reasoning in sequential recommendation from the ``slow thinking'' perspective. ManCAR~\cite{49_ManCAR} explicitly identifies the representation drift problem in latent reasoning, proposing a collaborative manifold built from interaction graphs to constrain reasoning trajectories. LCR-SER~\cite{50_LCRSER} performs latent cross-modal reasoning through dual-tower cross-attention iteration in a joint search-and-recommendation scenario, but this is essentially an information fusion mechanism rather than a hidden state feedback loop. The common limitation of the above methods is that they are based on the ID-embedding + dot-product discriminative ranking paradigm, a different technical route from LASAR's LLM decoder-only generative recommendation.

\textbf{Latent reasoning explorations in generative recommendation.} LatentR\textsuperscript{3}~\cite{51_LatentR3} is among the few works introducing latent reasoning into LLM-based recommendation; its LatentRATT module uses modified GRPO for two-stage training, but reasoning only extracts information from the LLM's final hidden state through single-layer attention. S\textsuperscript{2}GR~\cite{52_S2GR} inserts thinking tokens in generative retrieval and aligns them with SIDs, but this is essentially sequence insertion rather than recurrent iteration. Neither realizes the Coconut-style multi-step hidden state feedback loop in generative recommendation.

\textbf{Key distinctions from LASAR.} To our knowledge, LASAR is the first framework to perform complete latent reasoning through recurrent hidden-state feedback and adapt its reasoning depth to each sample in mainstream generative recommendation, with the following core features: (1) multi-step hidden state feedback loop with beam search; (2) explicit CoT semantic alignment (bidirectional KL) to prevent representation drift; (3) Policy Head + REINFORCE for sample-level adaptive step optimization. Systematic literature surveys~\cite{57_SurveyHowCanRSBenefitFromLLMs,58_SurveyTowardsLargeReasoningModels,59_TowardsReasoningAwareRS} indicate that latent reasoning research in recommender systems is still in its early stages. LASAR pioneers the integration of latent recurrent reasoning, LLM generative recommendation, and adaptive computation.
\section{Dataset Statistics}
\label{sec:app_dataset}

Table~\ref{tab:dataset} summarizes the statistics of the three Amazon review datasets used in our experiments.

\begin{table}[htbp]
\centering
\caption{Statistics of the three Amazon review datasets used in experiments.}
\label{tab:dataset}
\footnotesize
\setlength{\tabcolsep}{4pt}
\begin{tabular}{lrrrrr}
\toprule
Dataset & \#Users & \#Items & \#Inter. & Sparsity & \#Test \\
\midrule
Beauty & 22,363 & 12,101 & 176,139 & 99.935\% & 22,363 \\
Instruments & 24,772 & 9,922 & 74,316 & 99.970\% & 24,772 \\
Sports & 35,598 & 18,357 & 106,794 & 99.984\% & 35,598 \\
\bottomrule
\end{tabular}
\end{table}

\section{Evaluation Metrics}
\label{sec:app_metrics}

We adopt leave-one-out evaluation: the last interaction of each user is held out as the test item. For each user, the model generates a ranked list of candidates via beam search, and the following metrics are computed against the ground-truth next item.

\textbf{Hit Rate@$K$ (HR@$K$).} Measures whether the ground-truth item appears in the top-$K$ ranked list:
\begin{equation}
\text{HR@}K = \begin{cases} 1 & \text{if ground-truth item} \in \text{top-}K \\ 0 & \text{otherwise} \end{cases}
\end{equation}

\textbf{Normalized Discounted Cumulative Gain@$K$ (NDCG@$K$).} Assigns higher scores when the ground-truth item is ranked closer to the top. With a single ground-truth item at rank $r$ (where $r{=}1$ is the highest):
\begin{equation}
\text{NDCG@}K = \begin{cases} \frac{1}{\log_2(r+1)} & \text{if } r \leq K \\ 0 & \text{otherwise} \end{cases}
\end{equation}
where the ideal DCG (IDCG) is 1 since there is exactly one relevant item, making the normalization trivial.

We report both metrics at $K \in \{5, 10, 20\}$. Results are averaged over all test users. Significance is assessed with a one-sided paired bootstrap: for each dataset--metric, we resample test users with replacement (2000 replicates) and recompute the per-user metric for LASAR versus the strongest non-LASAR baseline on that dataset--metric, reporting the one-sided $p$-value of LASAR exceeding the baseline.

\section{Baseline Details}
\label{sec:app_baselines}

\emph{Traditional sequential models.} SASRec~\cite{28_SASRec} applies the self-attention mechanism to sequential recommendation, adaptively weighting historical items to capture long-term user preferences. GRU4Rec~\cite{30_GRU4Rec} is a pioneering session-based recommendation model that uses GRU to model user click sequences, representing the classical RNN-based approach.

\emph{LLM-based generative methods.} LC-Rec~\cite{36_LCRec} bridges collaborative filtering signals with LLM semantics through vector-quantized discrete item indices and multi-task alignment, enabling LLMs to perform end-to-end generative recommendation. MiniOneRec~\cite{37_MiniOneRec} is a compact generative recommender that directly generates item IDs autoregressively without additional reasoning. It shares the same base model and prompt format as LASAR, ensuring that any performance difference reflects the reasoning mechanism rather than model capacity or prompt design.

\emph{Latent reasoning.} ReaRec~\cite{46_ReaRec} performs multiple forward passes in continuous hidden space with a globally fixed step count, trained with progressive reasoning curriculum and contrastive alignment. We adopt its published implementation as a representative latent reasoning baseline. We also examined LatentR\textsuperscript{3}~\cite{51_LatentR3} and S\textsuperscript{2}GR~\cite{52_S2GR}. The released LatentR\textsuperscript{3} evaluation encountered OOM during beam search on even 80-GB NVIDIA A800 under our setting, while no public implementation or checkpoint was available for S\textsuperscript{2}GR as of the submission date.

\emph{Explicit CoT reasoning.} GREAM~\cite{41_GREAM} is a multi-component framework integrating (1) 19 collaborative-semantic alignment prompts for data augmentation, (2) structured multi-step CoT reasoning supervision, and (3) SRPO reinforcement learning post-training. We retain GREAM's core CoT reasoning chain while simplifying the 19 alignment prompts to MiniOneRec's prompt format, ensuring that all generative baselines share identical input--output templates and training data. GREAM's own ablation explores all combinations of inference mode (generating CoT text vs.\ answering directly) and training strategy (SFT vs.\ SFT+RL); we adopt its strongest configuration (CoT SFT + direct answer inference). We denote this variant as Explicit CoT$_{\text{GREAM}}$ in result tables.

\section{Engineering Implementation Details}
\label{sec:app_engineering}

Table~\ref{tab:implementation} lists the key hyperparameters used across all experiments. All generative methods share the same base model (Qwen3-0.6B, scaled to 1.7B in the main text), data split, SID encoding ($M{=}4$ tokens per item, 256 entries per codebook, up to $256^4$ possible SID sequences), and prompt format, so that performance differences reflect the reasoning mechanism rather than model capacity or data. A fixed random seed (42) is used for data sampling and data loading. Training uses AdamW with a cosine learning-rate schedule and a 0.08 warmup ratio; the two-stage SFT schedule (Stage~1 plain SID grounding, Stage~2 latent reasoning with KL alignment) and the RL configuration (GRPO with terminal KL and REINFORCE for step allocation) follow the main text. All experiments run on $8{\times}$L40 (48\,GB). Inference uses beam search with width 50, ranking candidates by joint token probability, and the bootstrap resampling procedure used for significance testing follows the setup in Appendix~\ref{sec:app_metrics}.

\begin{table}[htbp]
\centering
\caption{Implementation details and hyperparameters.}
\label{tab:implementation}
\begin{tabular}{ll}
\toprule
Parameter & Value \\
\midrule
Optimizer & AdamW \\
Batch size & 512 \\
LR scheduler & Cosine, warmup ratio 0.08 \\
Max sequence length & 512 \\
\midrule
Stage 1 epochs & 10 \\
Stage 1 learning rate & $5\times 10^{-4}$ \\
Stage 2 epochs & 20 \\
Stage 2 learning rate & $5\times 10^{-5}$ \\
Max latent steps ($N_{\text{max}}$) & 8 \\
$\alpha_{\text{align}}$ (align\_weight) & 0.1 \\
$\beta_{\text{policy}}$ (policy\_weight) & 0.1 \\
\midrule
RL learning rate & $1\times 10^{-5}$ \\
RL epochs & 1 \\
RL num\_generations ($G$) & 8 \\
RL $\beta$ (GRPO ref-KL coef.) & $1\times 10^{-3}$ \\
RL $\varepsilon$ (clip ratio) & 0.2 \\
RL $\gamma_{\text{RF}}$ (reinforce\_coef) & 0.01 \\
RL reinforce\_n\_penalty ($\lambda$) & \{0.0001, 0.0005, 0.001\} \\
RL $\gamma_{\text{KL}}$ (terminal align.\ coef.) & $1\times 10^{-5}$ \\
\midrule
Teacher CoT model & GPT-5 \\
Semantic segmentation & BAAI/bge-small-en-v1.5 \\
Beam search width & 50 \\
GPU & $8\times$ L40 (48GB) \\
\bottomrule
\end{tabular}
\end{table}

\paragraph{LoRA configuration (1.7B).} The 1.7B scaling experiments use LoRA with rank $r{=}16$, scaling factor $\alpha_{\mathrm{LoRA}}{=}32$, and dropout $0.05$, applied to all attention and MLP projection modules (\texttt{q\_proj, k\_proj, v\_proj, o\_proj, gate\_proj, up\_proj, down\_proj}).

\paragraph{SID item embeddings.} Following the Residual Quantization K-Means pipeline of prior generative recommenders~\cite{41_GREAM,61_HandbookSID}, each item's title and description are concatenated and encoded by Qwen3-4B-Instruct; the item embedding is the mean-pooled last-layer hidden state over non-padding token positions, computed offline. The \texttt{bge-small-en-v1.5} encoder is used only for semantic segmentation of teacher CoT traces.

\paragraph{REINFORCE hyperparameters.} The entropy coefficient is $\eta{=}0.01$, and the EMA baseline uses a decay factor of $0.99$, i.e., $b_{\text{EMA}} \leftarrow 0.99\,b_{\text{EMA}} + 0.01\,\bar{r}$. The full RL phase runs for one epoch over the training set; the backbone is optimized by GRPO and terminal KL, while the Policy Head is optimized by REINFORCE.

\subsection{Variable-Length $N$ Batch Processing}
\label{sec:app_batch}

LASAR processes a variable per-sample $N$, resulting in varying numbers of latent tokens within a batch. LASAR handles this through a unified padding-and-masking strategy (Figure~\ref{fig:batch_layout_app}) that requires no per-sample branching during the latent loop:
\begin{figure}[htbp]
  \centering
  \resizebox{0.9\linewidth}{!}{%
  \begin{tabular}{c | c c c | c c c c c | c}
  \toprule
  & \multicolumn{3}{c|}{Prompt (left-padded)} & \multicolumn{5}{c|}{Latent region} & \\
  Sample & $p_1$ & $p_2$ & $p_3$ & \texttt{<s>} & \texttt{<t>} & \texttt{<t>} & \texttt{<t>} & \texttt{<e>} & Answer \\
  \midrule
  A ($N{=}2$, short) & \texttt{PAD} & $x_1$ & $x_2$ & \texttt{<s>} & \texttt{<t>} & \texttt{<t>} & \texttt{PAD} & \texttt{<e>} & $Y_A$ \\
  B ($N{=}3$, long) & $x_1$ & $x_2$ & $x_3$ & \texttt{<s>} & \texttt{<t>} & \texttt{<t>} & \texttt{<t>} & \texttt{<e>} & $Y_B$ \\
  \midrule
  Attn (A) & $0$ & $1$ & $1$ & $1$ & $1$ & $1$ & $0$ & $1$ & $1$ \\
  Attn (B) & $1$ & $1$ & $1$ & $1$ & $1$ & $1$ & $1$ & $1$ & $1$ \\
  Loss mask & $0$ & $0$ & $0$ & $0$ & $0$ & $0$ & $0$ & $0$ & $1$ \\
  \bottomrule
  \end{tabular}%
  }
  \caption{Detailed batch layout for variable $N$: prompt left-padding aligns \texttt{<s>} positions, latent region is right-padded with masked attention, and loss is computed only on answer tokens.}
  \label{fig:batch_layout_app}
  \end{figure}

\begin{enumerate}
\item \textbf{Prompt left-padding}: Left-pad prompts with \texttt{pad\_token\_id} to the batch's maximum prompt length, aligning the first $\langle\text{thought}\rangle$ position across all samples.
\item \textbf{Per-sample latent insertion}: Insert $N_i$ copies of $\langle\text{thought}\rangle$ per sample after the prompt, producing naturally different latent region lengths.
\item \textbf{Latent region right-padding}: For samples with $N_i < \max(N)$, pad the remaining latent slots with \texttt{pad\_token\_id} (\emph{not} $\langle\text{thought}\rangle$) and set attention mask to 0. The latent loop iterates $\max(N)$ times uniformly; extra steps for short-$N$ samples operate on masked positions and produce hidden states that are ignored by subsequent attention.
\item \textbf{Sequence right-padding}: Use \texttt{pad\_sequence} to align total sequence lengths (prompt + latent region + answer) to the batch maximum.
\item \textbf{Loss masking}: The LM loss is computed only on answer tokens, and the alignment loss filters positions exceeding each sample's actual $N_i$ via a \texttt{valid\_mask}, so padded latent positions contribute zero gradient.
\end{enumerate}

\subsection{Prompt Format and CoT Reasoning Details}
\label{sec:app_prompt_cot}

All generative methods (LASAR, MiniOneRec, LC-Rec, and Explicit CoT$_{\text{GREAM}}$) share an identical prompt template following MiniOneRec~\cite{37_MiniOneRec}, so performance differences reflect the reasoning mechanism rather than prompt design. The template takes the form:

\smallskip
\noindent\fbox{\parbox{0.95\linewidth}{\small\texttt{Below is an instruction that describes a task, paired with an input that provides further context. Write a response that appropriately completes the request.}\\[2pt]
\texttt{\#\#\# Instruction:}\\
\texttt{Can you predict the next possible item that the user may expect?}\\[2pt]
\texttt{\#\#\# User Input:}\\
\texttt{The user has interacted with items <SID\_1>, <SID\_2>, ..., <SID\_L> in chronological order. Can you predict the next possible item that the user may expect?}\\[2pt]
\texttt{\#\#\# Response:}\\
\texttt{<SID\_target>}}}
\smallskip

\noindent where each \texttt{<SID\_i>} is a fixed-length four-token semantic ID (e.g., \texttt{<|a\_125|><|b\_109|><|c\_135|><|d\_125|>}), not a natural-language title. Items are listed in chronological order and separated by commas, following the MiniOneRec natural-language template; there is no dedicated reserved boundary token beyond this comma delimiter.

\paragraph{Teacher CoT Generation.}
The teacher CoT reasoning text is generated offline by GPT-5 (with \texttt{reasoning\_effort} set to \texttt{low}), producing one reasoning trace per training sample in a single pass; this cost is incurred only once, before training, and never at inference. Teacher traces are generated exclusively from training instances and their training-set targets; no validation or test targets are used for trace construction. Generated traces were spot-checked on a sampled subset to confirm overall quality; no per-trace manual filtering was applied. Following the reverse-reasoning distillation idea of GREAM~\cite{41_GREAM} (where the teacher is shown the ground-truth next item but instructed to reason \emph{as if} it were unknown, so the trace forms a plausible causal path from the interaction history toward a recommendation), our teacher prompt requests a four-step structure: \emph{(1) analyze the purchase history}, \emph{(2) build user preferences}, \emph{(3) uncover purchase intent}, and \emph{(4) recommend the item}. Steps~1--3 constitute the forward reasoning chain, whereas Step~4 states the recommended item itself.

\paragraph{Step Selection for Latent Alignment.}
LASAR aligns its latent reasoning steps only to Steps~1--3 of the teacher trace (analyze history, build preferences, uncover intent) and excludes Step~4 before segmentation. Latent alignment targets a plausible \emph{forward inferential trajectory} supported by the interaction history, rather than the answer representation itself. Steps~1--3 form the \emph{evidence $\rightarrow$ preference $\rightarrow$ intent} forward chain.

\paragraph{CoT Segmentation.}
The retained CoT text is segmented by a semantic splitter built on sentence embeddings from BAAI/bge-small-en-v1.5~\cite{64_bge_embedding}. The text is first split into sentences; a sliding window of \texttt{buffer\_size}${=}2$ adjacent sentences forms candidate chunks; each chunk is embedded with bge, and the cosine distance between consecutive chunk embeddings is computed. A breakpoint is inserted wherever this distance exceeds the 80th percentile of all consecutive distances in the trace (\texttt{breakpoint\_percentile\_threshold}${=}80$), so boundaries fall at the points of largest semantic shift. The number of segments $N$ is thus data-dependent, and directly serves as the Policy Head's per-sample supervision label during SFT.

\noindent\textbf{CoT Example.} Following GREAM's reverse-reasoning distillation, the teacher receives the user's purchase history together with the next item, and is instructed to reason \emph{as if} the next item were unknown. As in GREAM, the observed next item is treated as one of many plausible outcomes rather than a deterministic label, so the teacher constructs a plausible causal link from the history to a recommendation instead of replicating the observed item; the resulting trace is used for alignment supervision, with Steps~1--3 retained and Step~4 excluded. Below is an example from the Sports dataset (user 6729, history length 5).

\smallskip
\noindent\fbox{\parbox{0.95\linewidth}{\footnotesize
\textbf{Teacher input --- user purchase history (titles shown for brevity):}\\[3pt]
\textbf{Item 1:} 15 Pack of 16 Gram Threaded CO2 Cartridges.\\
\textbf{Item 2:} Cateye HL-EL135N Bicycle Head Light.\\
\textbf{Item 3:} Concept2 Model D Indoor Rowing Machine.\\
\textbf{Item 4:} Petzl Tikka XP 2 Headlamp.\\
\textbf{Item 5:} Delta Leonardo Single-Bike Rack.\\[3pt]
\textbf{Next Item:} Planet Bike Blinky Super Flash Tail Light.}}
\smallskip

\begin{framed}\footnotesize
\noindent\textbf{Teacher output --- full CoT (Steps 1--4 as originally generated):}\\[3pt]
\textbf{Step 1: Analyze the purchase history:}\\
- Strong cycling focus: CO2 cartridges (tire inflation), a bicycle headlight for visibility, and a wall-mounted bike rack for storage.\\
- Emphasis on safety/visibility and lighting: bicycle headlight and a high-performance headlamp (Petzl) for hands-free illumination.\\
- Preference for durable, high-quality, and practical gear with long runtimes, easy installation, and versatility (Cateye light, Petzl headlamp, Concept2 rower).\\
- Signs of active riding/commuting and home organization for the bike.\\
- Potential gaps: no rear tail light listed, no mention of reflective gear, no CO2 inflator head (though they may already own one), and no lock or saddle bag.\\[3pt]
\textbf{Step 2: Build user preferences:}\\
- Prioritizes safety and visibility while cycling (front light, lighting redundancy).\\
- Values long battery life, reliability, and compact, easy-mount designs.\\
- Favors reputable, performance-oriented brands.\\
- Seeks practical cycling accessories that complement daily/commuter use and home storage.\\[3pt]
\textbf{Step 3: Uncover purchase intent:}\\
- After adding a front headlight, a highly likely next step is completing the safety setup with a rear tail light for visibility to traffic.\\
- A rear blinker with wide side visibility, multiple flash modes, long runtime, and easy mounting aligns with their safety focus and gear choices.\\
- Brand consistency may be appealing given the existing CatEye headlight.\\[3pt]
\colorbox{black!15}{\parbox{0.93\linewidth}{\textbf{Step 4: Recommend the item:} \emph{(removed before alignment)}\\
- Title: CatEye Omni 5 Rear Bicycle Tail Light\\
- Description: The CatEye Omni 5 Rear Tail Light enhances on-road safety with five bright LEDs, 360-degree Omni-directional optics, and multiple modes (constant and flashing) for day and night visibility. Running up to 120 hours on two AAA batteries, it offers long-lasting performance, is weather-resistant, and installs tool-free with a secure FlexTight bracket on seatposts, seatstays, or bags. Compact and lightweight with a quick-release mount, it is an easy, reliable upgrade to complete a commuter-ready lighting setup.}}\end{framed}
\smallskip

\noindent Steps~1--3 are semantically segmented and encoded into alignment anchors; Step~4 is excluded before segmentation. In this trace, Step~1 identifies the absence of a rear tail light despite the presence of a front bicycle light, Step~2 abstracts the user's safety and visibility preference, and Step~3 infers the intent to complete the lighting setup with a rear blinker offering flash modes and side visibility, the same category and attributes as the observed next item. Aligning latent steps to this forward chain (evidence $\rightarrow$ preference $\rightarrow$ intent) provides a semantically relevant trajectory for next-item prediction, even though Step~4's specific pick (a CatEye tail light) is a plausible prediction rather than a replication of the observed item (a Planet Bike Blinky) and is discarded. The model is trained to generate the next item's SID via cross-entropy during the SFT phase. None of this CoT text is decoded at inference: the model aligns its latent hidden states to the Step~1--3 segments via bidirectional KL divergence during SFT, and the Policy Head learns to allocate reasoning depth in proportion to the number of segments.

\subsection{SFT Phase Alignment Implementation}

\paragraph{Anchor construction.}
Each CoT anchor is obtained in a \emph{single} forward pass over the full reconstructed CoT text (segments rejoined in order): we take the last-token hidden state at each segment boundary from the final Transformer layer. Each anchor is therefore contextualized by the preceding CoT segments, serving as a textual-reasoning reference that the latent trajectory aligns to. The CoT anchors are pre-computed offline and kept fixed; gradients are propagated only through the latent states.

During SFT, each latent step's hidden state must be aligned with the corresponding explicit CoT segment's hidden state:
\begin{enumerate}
\item \textbf{Semantic segment matching}: Use \texttt{cumsum} to compute each latent token's ordinal within the batch, vectorized-matching to the corresponding CoT segment.
\item \textbf{Boundary safety}: Mask padded latent positions (batch padding to $\max_i N_i$); each non-padding latent position is matched in order to its corresponding CoT segment, since $N_i$ equals the segment count by construction.
\item \textbf{Device consistency}: CoT embeddings are preloaded on CPU and dynamically converted to GPU dtype and device during training.
\end{enumerate}

\subsection{RL Phase Terminal Alignment}

\paragraph{What constrains the latent states during RL.}
During the RL phase the latent states receive gradients from two sources: (i) the GRPO objective, which scores answer-token log-probabilities and back-propagates through the latent forward pass into the base LM; and (ii) the terminal KL loss on the final latent step. The step-wise alignment used in SFT is not applied here because $N$ is sampled per instance, so variable-length reasoning chains no longer correspond one-to-one to the fixed set of CoT segments. The trajectory is anchored at both ends: it starts from the SFT-aligned policy and its final step is aligned by terminal KL, while GRPO shapes the reasoning in between through the recommendation reward.

RL-phase alignment targets only the \textbf{last} latent step of the reasoning chain:
\begin{enumerate}
\item \textbf{Natural latent embedding collection}: During Phase~2 training, hidden states from each step of the latent loop are automatically collected as a \texttt{latent\_embs} list (cached from the forward pass, with no extra forward passes needed).
\item \textbf{Dynamic step indexing}: \texttt{latent\_embs[$N$$-$1]} directly retrieves the last step's hidden state for bidirectional KL computation with the CoT final state.
\item \textbf{Alignment as direct loss, not reward}: If alignment were a reward component, GRPO's within-group advantage zero-mean property would cancel it out. Adding it as a direct loss to the total objective ensures stable gradient signals.
\end{enumerate}

\subsection{Reward Formulation}
\label{sec:app_reward}

For each prompt, $G$ candidates are generated via beam search. The exact match reward is:
\begin{equation}
r_{\text{rule}}^{(i)} = \begin{cases} 1 & \text{if } \hat{y}^{(i)} = y^* \\ 0 & \text{otherwise} \end{cases}
\end{equation}
where $i\in\{0,\dots,G-1\}$ indexes the candidate in the beam-searched group (0-indexed ranking position, $i=0$ for the top-ranked candidate), $\hat{y}^{(i)}$ is its generated SID, and $y^*$ is the ground-truth SID.

The within-group NDCG reward penalizes non-target candidates by their ranking position within the group. Let $w_i = -1/\log_2(i+2)$ (for $i=0,\dots,G-1$) be the position weight, which is negative with larger magnitude for higher-ranked candidates. If the ground-truth item appears among the $G$ candidates, non-target candidates receive:
\begin{equation}
r_{\text{NDCG}}^{(i)} = \begin{cases} 0 & \text{if } \hat{y}^{(i)} = y^* \\ \frac{w_i}{\sum_{j=0}^{G-1} |w_j|} & \text{if } \hat{y}^{(i)} \neq y^* \end{cases}
\end{equation}
If the ground-truth item is absent from all $G$ candidates, every candidate receives $r_{\text{NDCG}}^{(i)} = 0$. Since $r_{\text{rule}}$ is also $0$ for all candidates in this case, the group's rewards are all zero, so the normalized advantage $\hat{A}_i$ is zero and the group contributes no gradient.

\subsection{Inference Algorithm}
Algorithm~\ref{alg:inference} outlines LASAR inference: the Policy Head fixes a per-sample step count $N_i$; prompts are batch-aligned so the recurrent latent loop runs uniformly to $\max(N_i)$ with masked padding for shorter samples; and a trie-constrained beam search decodes the final item SID from the updated KV cache.

\begin{algorithm}[t]
\caption{LASAR Inference}
\label{alg:inference}
\begin{algorithmic}[1]
\State $N_i \gets \text{PolicyHead}(\text{prompt}_i)$ for each sample $i$
\State Insert \texttt{<s>\,<t>\,$\times$\,$N_i$\,<e>} after each prompt
\State Batch-align: left-pad prompts, right-pad latent region with \texttt{PAD} (mask=0)
\Statex
\State $h_0, \text{KV} \gets \text{LLM}(\text{prompt tokens})$ \Comment{Encode prompt}
\For{$t = 1$ to $\max(N_i)$} \Comment{Latent loop (equiv.\ to Eq.~3 of the main paper via KV cache)}
    \State $h_t \gets \text{LLM}(h_{t-1},\; \text{KV})$ \Comment{PAD steps ignored by mask}
\EndFor
\State Update KV cache with latent hidden states $\{h_t\}$ \Comment{Prepare for answer generation}
\Statex
\State Beam search from KV, constrained by item prefix tree \Comment{Decode the $M$-token SID}
\State \Return Top-K recommended items
\end{algorithmic}
\end{algorithm}

\subsection{Training Algorithm}
\label{sec:app_training}

Algorithm~\ref{alg:training} summarizes the full LASAR training pipeline: SID construction, offline anchor preparation, two-stage SFT, and RL post-training.

\begin{algorithm}[t]
\caption{LASAR Training}
\label{alg:training}
\begin{algorithmic}[1]
\State Construct item SIDs via Residual Quantization K-Means
\State Generate and segment teacher CoT traces for training samples
\State Encode CoT segments and cache fixed alignment anchors
\State \textbf{Stage 1: SID grounding}
\For{each Stage-1 minibatch}
    \State Optimize the answer-token cross-entropy loss
\EndFor
\State \textbf{Stage 2: Latent reasoning SFT}
\For{each Stage-2 minibatch}
    \State Set $N_i$ to the CoT segment count for each sample $i$
    \State Execute the masked variable-$N$ batch loop to $\max_i(N_i)$
    \State Compute SID generation, step-wise alignment, and Policy Head CE (warm-start) losses
    \State Update the backbone and Policy Head
\EndFor
\State \textbf{RL post-training}
\For{each prompt group}
    \State Sample $N \sim \pi_\phi(\cdot \mid h_0)$
    \State Generate $G$ candidate SIDs and compute group rewards
    \State Compute GRPO, terminal KL, and REINFORCE objectives
    \State Update the backbone and Policy Head using their respective objectives
\EndFor
\end{algorithmic}
\end{algorithm}

\subsubsection{Trie-Constrained Decoding}
\label{sec:app_trie}

Since SID tokens are hierarchical codes $(s_1, s_2, \dots, s_M)$ with each $s_j$ drawn from codebook $\mathcal{C}^{(j)}$, not every token combination corresponds to a valid item. Unconstrained beam search may produce invalid SID sequences that have no item in the catalog.

To ensure generation validity, we build a \textbf{prefix tree (trie)} over all item SIDs in the catalog. Each root-to-leaf path encodes exactly one item's SID sequence. During beam search, at each decoding step $j$, the trie is queried with the current prefix $(s_1, \dots, s_{j-1})$ to retrieve the set of valid next tokens $\mathcal{V}_j \subseteq \mathcal{C}^{(j)}$. A custom \texttt{ConstrainedLogitsProcessor} masks out all tokens outside $\mathcal{V}_j$ before the softmax, so every beam is guaranteed to trace a valid root-to-leaf path. This adds negligible overhead since trie lookups are $O(M)$ per step and the masking is a single tensor operation.

\begin{figure}[htbp]
  \centering
  \begin{tikzpicture}[
    level distance=14mm,
    sibling distance=12mm,
    every node/.style={draw, rounded corners, minimum size=6mm, inner sep=1.5pt, font=\small},
    edge from parent/.style={draw, -stealth, thick},
    level 1/.style={sibling distance=28mm},
    level 2/.style={sibling distance=14mm},
  ]
  \node {$\emptyset$}
    child {node {$s_1^1$}
      child {node {$s_2^1$}
        child {node[label={[font=\scriptsize,draw=none]below:Item A}] {$s_3^1$}}
        child {node[label={[font=\scriptsize,draw=none]below:Item B}] {$s_3^2$}}
      }
      child {node {$s_2^3$}
        child {node[label={[font=\scriptsize,draw=none]below:Item C}] {$s_3^1$}}
      }
    }
    child {node {$s_1^2$}
      child {node {$s_2^2$}
        child {node[label={[font=\scriptsize,draw=none]below:Item D}] {$s_3^3$}}
      }
      child {node {$s_2^4$}
        child {node[label={[font=\scriptsize,draw=none]below:Item E}] {$s_3^2$}}
        child {node[label={[font=\scriptsize,draw=none]below:Item F}] {$s_3^4$}}
      }
    };
  \end{tikzpicture}
  \caption{Trie over item SIDs ($M{=}3$, $|\mathcal{C}^{(j)}|{=}4$). Each root-to-leaf path is a valid item. At step $j$, only children of the current prefix are allowed, preventing invalid sequences.}
  \label{fig:trie_example}
\end{figure}
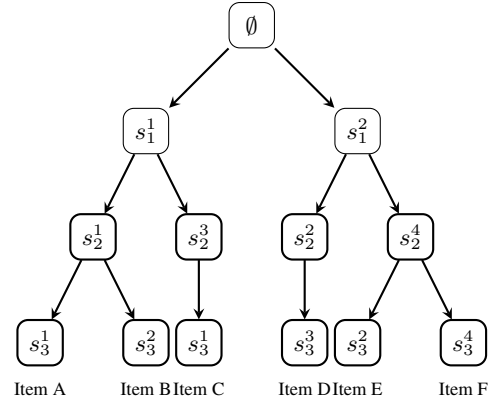

Figure~\ref{fig:trie_example} illustrates this process with a simplified example where $M{=}3$ and each codebook has 4 tokens.

\section{SFT Labels and Sample Complexity}
\label{sec:app_sft_labels}

In the SFT phase, the Policy Head's supervision labels come from the Teacher CoT semantic segment count. Table~\ref{tab:teacher_cot} shows that samples with more segments consistently have longer histories and higher category diversity: interaction histories that span more diverse categories and longer sequences yield more CoT segments and thus a larger predicted depth. The Policy Head therefore learns a difficulty-aware allocation, assigning more latent steps to more complex user patterns.

\paragraph{Fixed-$N$ comparison protocol.}
The fixed-$N \in \{1,4,8\}$ curves in the Force-$N$ analysis (main text, RQ3) are produced from the trained LASAR checkpoint: at inference we disable the Policy Head and set every sample to a constant number of latent steps. This analysis asks whether the depth the Policy Head assigns to each sample actually helps, so we vary only the allocation while leaving the model itself unchanged. Because every learned parameter is held fixed, the performance gap reflects how reasoning depth is distributed across samples.

\begin{table}[htbp]
  \centering
  \caption{Teacher CoT segment count vs.\ sample complexity on Sports.}
  \label{tab:teacher_cot}
  \begin{tabular}{lccc}
  \toprule
  Segments & Prop.\ & Hist.\ Len.\ & Cat.\ Div.\ \\
  \midrule
  3 segments & 83.4\% & 6.9 & 6.2 \\
  4 segments & 16.3\% & \textbf{8.5} & \textbf{7.5} \\
  5 segments & 0.2\% & \textbf{9.8} & \textbf{8.8} \\
  \midrule
  \textbf{Trend} & & $\uparrow$ & $\uparrow$ \\
  \bottomrule
  \end{tabular}
\end{table}

\section{Alignment Ablation on Sports and Instruments}
\label{sec:app_sft_cross}

Table~\ref{tab:sft_cross} extends the alignment ablation (Table~2) to Sports and Instruments. KL alignment consistently yields the best results across both datasets.

\begin{table}[htbp]
\centering
\caption{Alignment ablation on Sports and Instruments. Best in \textbf{bold}.}
\label{tab:sft_cross}
\resizebox{\linewidth}{!}{%
\begin{tabular}{llcccc}
\toprule
& Model & N@5 & N@10 & HR@5 & HR@10 \\
\midrule
\multirow{3}{*}{Sports}
& Pure SFT & 0.0105 & 0.0136 & 0.0163 & 0.0260 \\
& + Latent (w/o align.) & 0.0104 & 0.0139 & 0.0168 & 0.0276 \\
& \textbf{+ KL Alignment} & \textbf{0.0109} & \textbf{0.0143} & \textbf{0.0174} & \textbf{0.0281} \\
\midrule
\multirow{3}{*}{Instruments}
& Pure SFT & 0.0580 & 0.0634 & 0.0720 & 0.0888 \\
& + Latent (w/o align.) & 0.0517 & 0.0576 & 0.0693 & 0.0878 \\
& \textbf{+ KL Alignment} & \textbf{0.0586} & \textbf{0.0643} & \textbf{0.0731} & \textbf{0.0910} \\

\bottomrule
\end{tabular}%
}
\end{table}

KL alignment consistently yields the best results across both datasets, improving NDCG@10 over Pure SFT. The unaligned variant shows inconsistent behavior, with slight gains on some Sports metrics but notable degradation on Instruments ($-$9.1\% NDCG@10). This confirms that KL-based alignment is not dataset-specific: it provides stable improvement, while unaligned latent reasoning does not reliably benefit recommendation.

\end{document}